\pdfoutput=1  
\documentclass[11pt, letterpaper]{article}

\usepackage[margin=1in]{geometry}

\usepackage[T1]{fontenc}
\usepackage{mathptmx}  

\usepackage{setspace}
\usepackage{titlesec}
\titleformat{\section}{\normalfont\large\bfseries}{\thesection.}{0.5em}{}
\titleformat{\subsection}{\normalfont\bfseries}{\thesubsection.}{0.5em}{}
\titleformat{\subsubsection}{\normalfont\itshape}{\thesubsubsection.}{0.5em}{}

\usepackage{graphicx}
\graphicspath{
  {figures/}
  {../outputs/figures/}
}

\usepackage{booktabs}
\usepackage{tabularx}

\usepackage{amsmath}
\usepackage{amssymb}

\usepackage[dvipsnames]{xcolor}
\usepackage[hidelinks]{hyperref}
\hypersetup{
  colorlinks=true,
  linkcolor=MidnightBlue,
  citecolor=MidnightBlue,
  urlcolor=MidnightBlue,
}

\usepackage[numbers,sort&compress,super]{natbib}
\usepackage{enumitem}        
\setlist{nosep,leftmargin=*} 
\usepackage{xspace}

\newcommand{\ie}{i.e.\xspace}
\newcommand{\eg}{e.g.\xspace}

\newcommand{\mz}{\textit{m/z}\xspace}
\newcommand{\rt}{RT\xspace}

\begin{document}

\title{The Language of Elution: Autoregressive Prediction of the\\Next Feature in Untargeted LC-HRMS Lipidomics}

\author{
  Dayanjan S.\ Wijesinghe\textsuperscript{1,*}
}

\date{}  

\maketitle

\noindent
\textsuperscript{1}Department of Pharmacotherapy and Outcomes Sciences,
Virginia Commonwealth University School of Pharmacy, Richmond, VA 23298, USA \\
\textsuperscript{*}Corresponding author: Dayanjan S.\ Wijesinghe,
\href{mailto:wijesingheds@vcu.edu}{wijesingheds@vcu.edu}\\
ORCID: \href{https://orcid.org/0000-0002-2124-5109}{0000-0002-2124-5109}

\vspace{1em}


\begin{abstract}
\noindent
Untargeted liquid chromatography--high-resolution mass spectrometry (LC-HRMS)
routinely detects thousands of molecular features per sample, yet only
2--20\% receive confident structural annotations. A root cause of this
``dark metabolome'' is that tandem mass spectrometry (MS/MS) acquisition
remains reactive: instruments select precursor ions after they appear, with no
foreknowledge of what will elute next. Here we reframe chromatographic elution
as an autoregressive sequence prediction task. Because reversed-phase elution
order is governed by hydrophobicity, successive features are not independent
draws but elements of a physically constrained sequence---analogous to tokens
in natural language. We discretize the mass-to-charge (\mz) axis into 110 bins
and train long short-term memory (LSTM) and Transformer models to predict the
next eluting \mz bin from five per-token, annotation-free input features: \mz bin, mass defect,
retention-time gap, ionization polarity, and intensity rank. Trained on 15,242
consensus features from four clinical lipidomics cohorts (342 human plasma
samples, SCIEX TripleTOF 6600+, Waters CSH C18), the LSTM achieves 98.4\%
top-1 accuracy (99.99\% top-5; mean absolute error = 3.6~Da) and the Transformer
achieves 98.0\% top-1 accuracy (99.88\% top-5; MAE = 4~Da). Ablation analysis reveals that autoregressive
sequence context accounts for 55.5 percentage points of accuracy; no individual
input feature contributes more than 0.2~pp, establishing that the sequential
pattern---not molecular properties---drives prediction. Cross-platform
validation on an independent Agilent 6530 dataset using the same
chromatographic method yields comparable performance (retention-time
correlation $r = 0.999$), whereas datasets with different column chemistry
(5.1\% top-1) or different polarity acquisition mode (2.6\% top-1, same
instrument and column) fail catastrophically, confirming that models are
specific to both chromatographic method and acquisition mode. However, full
fine-tuning on as few as two to five quality-control injections recovers held-out
analytical accuracy from 2.6\% to nearly 50\% top-1 (and to 99.6\% on held-out
QC), showing that cross-condition deployment is achievable with minimal
calibration. A QC warm-up
experiment reveals a null result: priming hidden states with quality-control
injections confers no benefit over cold-start inference, indicating that the
model captures elution logic from the sequence alone. Applying dual
mass-plus-retention-time filtering to model predictions yields 168 putative
annotations for previously unannotated features, illustrating potential for
dark-lipidome characterization. These results establish that chromatographic
elution sequences are highly predictable and lay the groundwork for predictive
MS/MS acquisition strategies that could substantially improve annotation
coverage in untargeted metabolomics.

\vspace{0.5em}
\end{abstract}

\clearpage


\section{Introduction}

Lipids are among the most abundant and structurally diverse molecules in
biology: a single mammalian cell contains thousands of distinct lipid species
spanning categories as varied as fatty acyls, glycerolipids,
glycerophospholipids, sphingolipids, sterols, and
prenols.\cite{quehenberger2010lipidome,harayama2018diversity}
Collectively termed the \textit{lipidome}, this ensemble underpins functions
that reach into nearly every aspect of physiology. Lipids are the principal
structural components of cellular membranes, where their composition sets the
physical properties of the bilayer and modulates the folding, trafficking, and
signaling of embedded membrane proteins;\cite{harayama2018diversity}
cholesterol- and sphingolipid-enriched membrane microdomains (``lipid rafts'')
further organize receptors and signaling complexes laterally within the
membrane.\cite{sezgin2017rafts}
Beyond structure, neutral lipids packaged into lipid droplets serve as the
cell's primary energy reservoir and a buffer against lipotoxic
stress,\cite{olzmann2019lipiddroplets} while a large repertoire of bioactive
lipids---eicosanoids, sphingolipids, and phosphoinositides among them---act as
potent signaling molecules that govern inflammation, proliferation, apoptosis,
and metabolism.\cite{dennis2015eicosanoid,hannun2018sphingolipids,wymann2008lipidsignalling}
Reflecting this centrality, dysregulated lipid metabolism is a driver of major
human disease, from the lipid-laden plaques of atherosclerosis to a broad
spectrum of metabolic and inflammatory
disorders.\cite{libby2019atherosclerosis,lydic2018lipidomecomplexity}
Characterizing the lipidome at the level of individual molecular species
therefore carries substantial diagnostic and mechanistic value, but it demands
an analytical platform that can resolve thousands of structurally similar
species across a wide dynamic range. Liquid chromatography coupled to
high-resolution tandem mass spectrometry (LC-HRMS) has become the method of
choice for lipid discovery and quantitation, pairing chromatographic separation
of isobaric and isomeric species with the mass accuracy and fragmentation
needed for confident structural assignment.\cite{wei2019lipidomes,cajka2017validating}

Untargeted LC-HRMS can detect thousands of molecular features in a single
biological sample,
making it a cornerstone of metabolomics and lipidomics
discovery.\cite{patti2012apogee,fiehn2014metabolomicsinsights}
Yet the vast majority of these features remain structurally
uncharacterized. Across large-scale studies, only 2--20\% of detected features
receive confident annotations at Schymanski confidence
level~2 or better,\cite{schymanski2014confidence,natcomm2022annotation}
leaving the remainder consigned to the so-called ``dark
metabolome.''\cite{daSilva2015darkMatter,monge2019darkMolecules}
This annotation gap is not merely an academic inconvenience: unannotated
features frequently carry biological significance. In a study of mild
cognitive impairment, 68\% of the statistically significant features lacked
structural identification;\cite{hajjar2020mci} and systematic formula-level
analysis of untargeted metabolomes has uncovered previously unrecognized
metabolites hidden among unknown features.\cite{chen2021netid}
The dark metabolome thus represents both the largest reservoir of untapped
biological information and the most pressing bottleneck in untargeted
workflows.

The field has responded with increasingly sophisticated computational tools
for post-acquisition annotation. Spectral library matching platforms such as
the Global Natural Products Social Molecular Networking
(GNPS)\cite{Wang2016gnps} and MassBank\cite{Horai2010massbank} enable
community-scale spectral comparison, while machine learning approaches---SIRIUS
for molecular formula prediction,\cite{Duhrkop2019sirius} CANOPUS for compound
class assignment,\cite{Duhrkop2021canopus} MS2DeepScore for learned spectral
similarity,\cite{Huber2021ms2deepscore} and CFM-ID for \textit{in silico}
fragmentation\cite{Allen2014cfmid}---have extended annotation reach beyond
reference libraries. In parallel, quantitative structure--retention
relationship (QSRR) models now predict chromatographic retention times from
molecular structure with cross-system accuracy on the order of
30--60~seconds,\cite{bonini2020retip,schmid2024rttransformer,witting2020rtprediction}
and retention-order-aware scoring has been shown to improve structural
annotation by up to 66.1\%.\cite{bach2022lcms2struct,sun2025roasmi}
Despite this progress, all of these approaches share a fundamental
limitation: they operate \textit{post hoc}, annotating features only after
the data have been collected.

The root cause of poor annotation coverage is not solely computational but
also instrumental. Current tandem mass spectrometry (MS/MS) acquisition
strategies are inherently reactive. In data-dependent acquisition (DDA), the
instrument selects the top-$N$ most intense ions for fragmentation in real
time, a stochastic process that yields MS/MS coverage of only
5--18\% of detected features per
injection.\cite{defossez2023dda,elboudlali2025comparison}
Data-independent acquisition (DIA) fragments all ions within broad isolation
windows but produces chimeric spectra that complicate
deconvolution.\cite{elboudlali2025comparison}
Iterative exclusion strategies such as AcquireX improve cumulative coverage
across repeated injections but remain reactive within each
run.\cite{cooper2024acquirex}
Software-defined acquisition controllers---ViMMS with reinforcement
learning,\cite{davies2021vimms} FLASHIda with machine-learning quality
filters\cite{jeong2022flashida}---optimize \textit{which} ions to fragment
once detected but do not anticipate what will appear next.
Even MaxQuant.\allowbreak{}Live, the closest approach to predictive scheduling, requires a
predefined target list and therefore cannot accommodate unknown
compounds.\cite{wichmann2019maxquantlive}
In untargeted metabolomics specifically, real-time library-search-driven
controllers such as Met-IQ trigger MS\textsuperscript{n} decisions from
spectra as they are acquired,\cite{bills2022metiq} and dynamic
data-independent acquisition adjusts isolation windows during the gradient
through real-time retrospective alignment.\cite{heil2023dynamicdia}
These too remain reactive, responding to ions or windows at the current
retention time rather than forecasting what is still to come.
To our knowledge, no existing system pre-configures MS/MS parameters for
compounds that have not yet eluted based on a prediction of what is coming.

The physical chemistry of reversed-phase liquid chromatography (RP-LC) suggests
that such prediction should be feasible. Elution order in RP-LC is governed
primarily by analyte hydrophobicity: compounds partition between the mobile and
stationary phases according to their lipophilicity, with more hydrophobic
species eluting later.\cite{white2022ecn,vankova2022retention}
For lipids, this relationship is well described by the equivalent carbon number
(ECN) model, where retention increases with acyl-chain length and decreases
with degree of unsaturation; head-group class then adds systematic offsets to
this hydrophobicity-driven order.\cite{white2022ecn}
Crucially, elution \textit{order} is more conserved across instruments and
laboratories than absolute retention time, as demonstrated by inter-laboratory
retention-order databases.\cite{bach2018retention,stanstrup2015predret}
Within a single analytical batch, retention-time reproducibility is
typically better than 2\% RSD.\cite{gonzalezdominguez2024qcomics}
In our own training data spanning four clinical cohorts analyzed on the same
platform over multiple years, cross-cohort retention-time correlation is
$r > 0.999$ with a mean absolute error of 4.5~seconds, and mass defect
(the fractional component of \mz) alone predicts retention time with
$r = 0.899$. These observations indicate that the chromatographic elution
sequence is not stochastic but physically constrained and therefore, in
principle, predictable.

If elution order is governed by deterministic physics, then the sequence of
features eluting from a chromatographic column can be treated as an ordered
``language'' amenable to the same autoregressive modeling frameworks that
underpin modern natural language processing. In autoregressive next-token
prediction, a model estimates the probability distribution over the next
element in a sequence conditioned on all preceding
elements---the foundation of large language
models.\cite{vaswani2017attention}
This paradigm has already proven effective on ordered molecular data:
protein language models predict amino-acid sequences with evolutionary
fidelity,\cite{rives2021esm} chemical reaction prediction treats reagents
and products as token sequences,\cite{schwaller2019molecular} and
autoregressive forecasting of analytical instrument signals has been
demonstrated in materials
science.\cite{lewis2022eelstm}
Recent self-supervised foundation models for mass spectrometry have begun to
incorporate retention-related information: DreaMS uses chromatographic
retention-order prediction as a pre-training objective to learn spectral
representations from 700~million MS/MS spectra,\cite{bushuiev2025dreams}
and LSM1-MS2 employs masked peak reconstruction for spectral
embedding.\cite{asher2024lsm1ms2}
However, these models learn \textit{representations of individual spectra}
using retention order as a self-supervised signal; they do not model the
elution stream itself as a generative sequence. Sequence and attention-based
architectures have likewise been applied to LC-MS data---for instance,
treating the chromatogram as a multivariate time series for peak
detection\cite{xu2020peakdetection}---but to segment or embed existing signal
rather than to generate the upcoming feature stream. To our knowledge, no prior
work has formulated untargeted LC-MS elution as an autoregressive
next-feature prediction problem.

\begin{figure}[!tb]
\centering
\includegraphics[width=\textwidth]{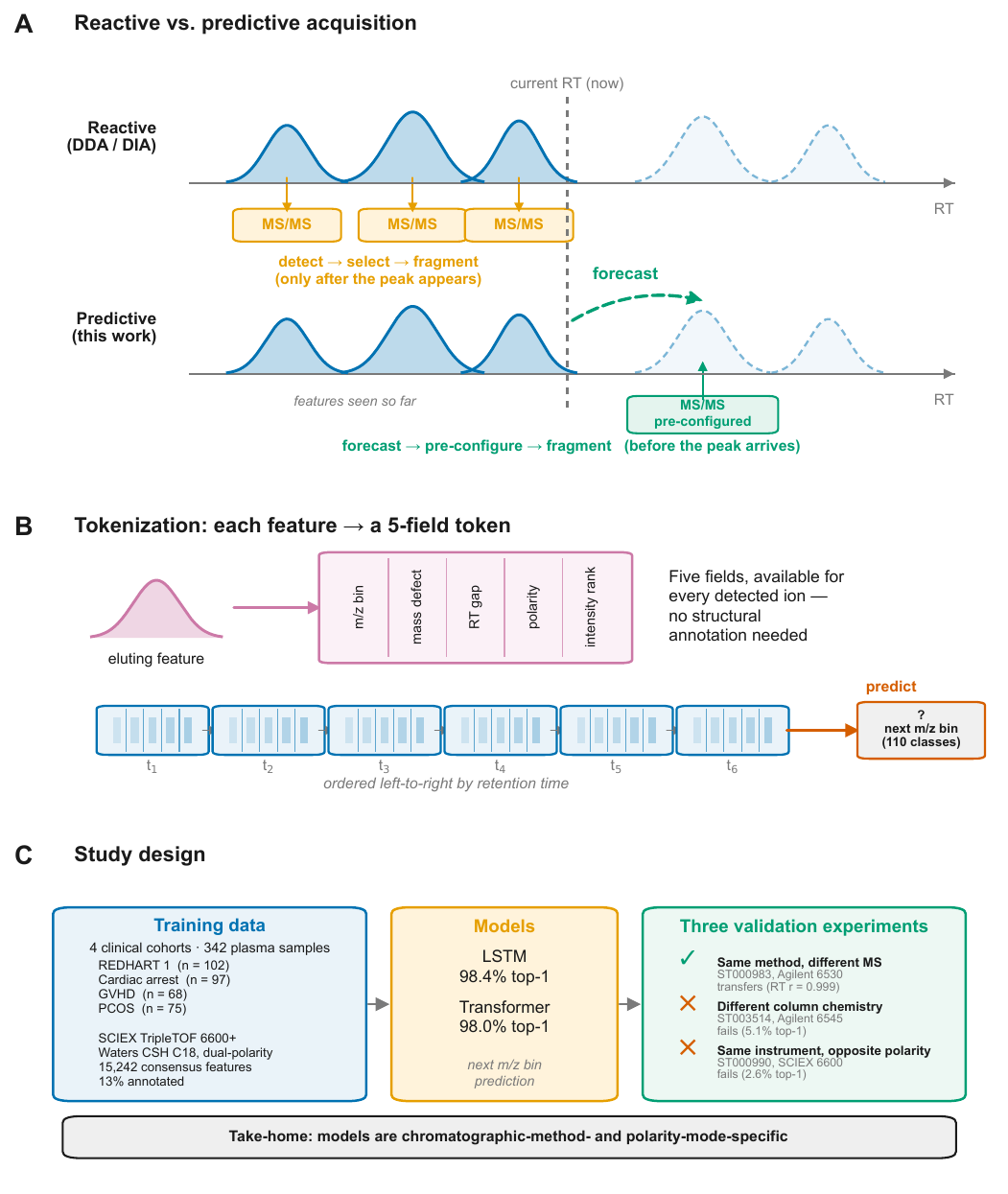}
\caption{\textbf{Study overview.}
\textbf{(A)}~Current LC-MS/MS acquisition is reactive: the instrument detects ions, then
selects precursors for fragmentation. Predictive acquisition would forecast upcoming
features before they elute, enabling pre-configured MS/MS parameters.
\textbf{(B)}~Tokenization scheme. Each detected feature is encoded as a composite token
with five input fields derived from the feature table, none of which require structural
annotation. Features are ordered by retention time within each sample to form an elution
sequence.
\textbf{(C)}~Study design. Models were trained on 15{,}242 consensus features from four
clinical lipidomics cohorts (342 samples) and validated across three external datasets
that systematically vary instrument platform and chromatographic method.}
\label{fig:overview}
\end{figure}

Here we introduce this formulation and demonstrate that the next eluting
feature in an LC-HRMS run can be forecast with high accuracy. Our approach
differs from prior retention-aware methods in both task structure and
operational goal. Structure-based QSRR models predict \textit{when} a known
molecule will elute; retention-order-aware annotation tools use pairwise
chromatographic constraints to improve \textit{post hoc} identification;
real-time acquisition frameworks reprioritize \textit{already detected} ions
or track predefined targets; and foundation models such as DreaMS use
retention order to learn \textit{spectral representations}. By contrast, we
model the within-run elution stream as an autoregressive process, asking
whether prior features contain sufficient information to forecast those that
follow---using only observed feature attributes, with no structural, spectral,
or candidate-compound input, over the predominantly unannotated detected
features. This is a distinct predictive task from structure-based retention
prediction, and one with implications for anticipatory MS/MS acquisition.

Specifically, we discretize the \mz axis into 110 bins and represent each
eluting feature as a token characterized by five input features: \mz bin,
mass defect, retention-time gap, ionization polarity, and intensity rank.
Trained on 15,242 consensus features from four clinical lipidomics cohorts
(342 human plasma samples), a long short-term memory (LSTM) network achieves
98.4\% top-1 accuracy on held-out test data, and a Transformer
achieves 98.0\%. We validate cross-platform generalization, showing that models
are chromatographic-method-specific rather than instrument-specific
($r = 0.999$ for matched chromatography; 5.1\% accuracy for mismatched
column chemistry). A controlled QC warm-up experiment yields a null result,
demonstrating that the model captures elution logic from the feature
sequence alone without requiring hidden-state conditioning. Finally, we apply
the trained model to the unannotated portion of the feature table, generating
168 putative lipid annotations via dual mass-plus-retention-time filtering
against reference databases. Together, these results establish chromatographic
elution as a highly predictable sequence and lay the groundwork for
predictive acquisition strategies in untargeted metabolomics.


\section{Methods}

\subsection{Datasets}

Four clinical lipidomics cohorts served as training and development data
(Table~\ref{tab:datasets}). All cohorts comprised human plasma samples analyzed
on a SCIEX TripleTOF 6600+ quadrupole time-of-flight mass spectrometer coupled
to a Waters Acquity UPLC with a CSH C18 column (100~mm $\times$ 2.1~mm,
1.7~$\mu$m), using both positive and negative electrospray ionization (ESI).
Raw data were processed with MS-DIAL\cite{tsugawa2015msdial} using
the LipidBlast (2017) in-silico spectral library for annotation. The four
cohorts---REDHART~1 (heart failure),\cite{vantassell2017redhart1,contaifer2019hf} Cardiac
Arrest, Graft-versus-Host Disease (GVHD),\cite{contaifer2019gvhd} and
Polycystic Ovary Syndrome (PCOS)---span distinct disease populations but share
identical analytical methodology, enabling assessment of cross-cohort
generalizability. Together, these cohorts yielded 15{,}242 consensus features
across 342 samples (308 analytical and 34 quality control [QC]), of which
1{,}976 (13\%) received structural annotations and 13{,}266 (87\%) remained
unannotated.

\begin{table}[htbp]
\centering
\small
\caption{Summary of training and validation datasets.}
\label{tab:datasets}
\begin{tabularx}{\textwidth}{lXrrr}
\toprule
Dataset & Instrument / Column & Features & Samples & Annotated \\
\midrule
Cardiac Arrest & SCIEX 6600+ / CSH C18 & 3{,}173 & 97 (87 + 10 QC) & 409 (12.9\%) \\
GVHD & SCIEX 6600+ / CSH C18 & 3{,}720 & 68 (62 + 6 QC) & 578 (15.5\%) \\
PCOS & SCIEX 6600+ / CSH C18 & 4{,}302 & 75 (67 + 8 QC) & 459 (10.7\%) \\
REDHART~1$^{b}$ & SCIEX 6600+ / CSH C18 & 4{,}047 & 102 (92 + 10 QC) & 530 (13.1\%) \\
\midrule
\textit{Training total} & & \textit{15{,}242} & \textit{342} & \textit{1{,}976 (13\%)} \\
\midrule
ST003514 (external) & Agilent 6545 QTOF / different C18 & 596 & 51 & 596 (100\%) \\
ST000983 (cross-platform) & Agilent 6530 QTOF / CSH C18 & 242$^{a}$ & 155 & 242 (100\%) \\
ST000990 (raw data) & SCIEX 6600 / CSH C18 (ESI+ only) & 3{,}604 & 153 & 307 (8.5\%) \\
\bottomrule
\end{tabularx}
\smallskip
\raggedright\footnotesize{$^{a}$Lipids matched to training cohorts by normalized annotated lipid name. $^{b}$REDHART~1 samples were drawn longitudinally (baseline and 3-month) from 57 patients; the 92 analytical injections therefore exceed the patient count.}
\end{table}

Three external datasets were used for validation. ST003514\cite{martinez2024srm1950}
from the Metabolomics Workbench comprises NIST SRM~1950 human plasma\cite{bowden2017nist}
analyzed on an Agilent 6545 QTOF with a different C18 column, providing a test
of generalization to mismatched chromatography. ST000983, from the Cajka and
Fiehn nine-platform comparison study\cite{cajka2017validating}, was acquired on an Agilent 6530 QTOF with
the same CSH C18 chromatographic method, enabling isolation of instrument
effects from chromatographic effects. ST000990, from the same nine-platform
study\cite{cajka2017validating}, was acquired on a SCIEX TripleTOF 6600 (same instrument family as
training data) with CSH C18 chromatography; raw .wiff files were reprocessed
with MS-DIAL to test generalization to independently processed data from a
near-identical platform.

\subsection{Ethics Approval and Informed Consent}

All four clinical lipidomics cohorts were derived from banked human plasma
collected in studies conducted at Virginia Commonwealth University (VCU) in
accordance with the Declaration of Helsinki and approved by the VCU
Institutional Review Board (IRB); all participants provided written informed
consent prior to enrollment. The heart-failure cohort (REDHART~1) was studied
under VCU IRB protocol HM15339,\cite{contaifer2019hf} with participants drawn
from the REDHART trial of interleukin-1 blockade in recently decompensated
systolic heart failure (ClinicalTrials.gov NCT01936909).\cite{vantassell2017redhart1}
The cardiac-arrest cohort was collected under VCU IRB protocol HM15326
(ClinicalTrials.gov NCT01944605). The graft-versus-host disease (GVHD) cohort
was enrolled in a VCU IRB--approved observational pilot study (VCU IRB Panel~A;
45~CFR~46.108(b), 46.109(e), and 46.110; approval date 27~March~2019).\cite{contaifer2019gvhd}
The polycystic ovary syndrome (PCOS) cohort was studied under VCU IRB protocol
10313. The present work is a secondary computational analysis of de-identified,
previously generated feature tables; it produced no new human-subjects data and
accessed no identifiable private information.

\subsection{Feature Table Construction}

MS-DIAL aligned features across all injections within each cohort to produce
consensus feature tables, assigning a single retention time per feature based on
alignment across samples. Each feature was characterized by \mz,
retention time, per-sample intensity, ionization polarity, and---for the 13\%
with library matches---lipid class and InChI Key. The four cohort-level feature
tables were combined into a unified dataset of 15{,}242 unique features.
Annotation rates were consistent across cohorts (10.7--15.5\%), suggesting the
annotation bottleneck reflects reference library coverage (LipidBlast) rather
than sample-specific factors.

\subsection{Tokenization}

We represented each detected feature as a composite token with five input
fields, selected based on systematic evaluation of all available features for
their contribution to retention time prediction. The tokenization was designed
to operate on 100\% of features, including the 87\% that lack structural
annotation.

\begin{enumerate}[leftmargin=2em]
\item \textbf{\mz bin} (110 occupied bins, 10~Da width, spanning 100--1{,}200~Da; the model's output layer has dimension 120, indexing \mz from 0, with the sub-100~Da bins never populated).
  The \mz value encodes both molecular weight, which scales with acyl chain
  length, and head group mass, which differs systematically across lipid classes
  (\eg phosphocholine = 184~Da, phosphoethanolamine = 141~Da). For unannotated
  features, \mz serves as the primary proxy for both hydrophobicity and
  structural class.

\item \textbf{Mass defect bin} (20 equal-width bins across $[0, 1)$). Mass
  defect---the fractional component of \mz---reflects hydrogen content, as each
  hydrogen atom adds 0.00783~Da above integer mass. More saturated acyl chains
  carry more hydrogens, yielding higher mass defect and later elution. Among
  annotated features, mass defect was the strongest univariate predictor of
  retention time ($r = 0.899$, $p < 10^{-300}$), capturing the combined effect
  of chain length and unsaturation in a single annotation-independent quantity.

\item \textbf{Retention time gap} (7 bins: co-elute $<$0.1~s, 0.1--0.5~s,
  0.5--1~s, 1--2~s, 2--5~s, 5--15~s, $>$15~s). The gap from the previous
  feature in the elution sequence encodes local feature density and
  chromatographic pace, providing temporal context that static features cannot
  capture.

\item \textbf{Ionization polarity} (3 levels: positive, negative, unknown).
  ESI mode is confounded with compound class---triglycerides and cholesteryl
  esters ionize preferentially in positive mode, while fatty acids favor
  negative mode---and therefore serves as a weak proxy for structural class.

\item \textbf{Intensity rank} (5 bins: top 1\%, 5\%, 20\%, 50\%, bottom
  50\%). Within-sample abundance percentile normalizes across the wide dynamic
  range of untargeted lipidomics ($10^2$--$10^7$). Although intensity is
  uncorrelated with retention time ($r = 0.066$), it provides contextual
  information about feature detectability relevant to acquisition scheduling.
\end{enumerate}

Each unique combination of these five fields (plus lipid class where annotated)
defines a composite token; the resulting composite vocabulary comprises 7{,}819
entries. The neural models do not consume composite-token IDs directly---they
embed the five fields separately and sum the embeddings (see Neural
Architectures)---so this vocabulary characterizes sequence diversity rather than
serving as the model input space. Sequences were constructed by sorting all detected features within each
sample by retention time, yielding a mean of ${\sim}3{,}717$ tokens per
sample. The prediction target was the \mz bin of the next feature in the
sequence (110 classes).

Several candidate features were evaluated and excluded. Adduct type showed no
significant residual contribution to retention time after controlling for
equivalent carbon number and lipid class ($p > 0.03$ for all adducts). Lipid
class, while the dominant retention time predictor among annotated features
($R^2 = 0.930$), was unavailable for 87\% of features and was therefore not
included as a primary token field; its information is captured implicitly by \mz
and mass defect.

\subsection{Train/Validation/Test Splitting}

Samples were split into training (242 samples), validation (50 samples), and
test (50 samples) sets using stratified random sampling by cohort to ensure
proportional representation of all four clinical populations. Splitting was
performed at the sample level rather than the feature level: the same molecular
feature could appear in both training and test samples, but the sequence context
in which it appeared differed. This design reflects the intended deployment
scenario, where the model encounters familiar molecular species in novel
elution contexts. The sliding-window procedure (context length = 64 tokens)
generated 883{,}696 training examples, 182{,}852 validation examples, and
182{,}902 test examples. The neural models were accordingly evaluated on these
182{,}902 fully-contexted test positions, whereas the baseline models---which do
not require a 64-token context window---were evaluated on all 186{,}052
predictable positions of the \emph{same} 50-sample held-out test set. The two
counts differ only by the 63 leading context tokens per test sample that the
sliding-window neural evaluation excludes; because accuracy is uniform across
sequence position (Figure~\ref{fig:position_accuracy}), this difference does not
materially affect any reported metric.

\subsection{Baseline Models}

Eight baselines spanning random, frequency-based, Markov, and linear
extrapolation strategies were evaluated on next \mz bin prediction (110
classes):

\begin{enumerate}[leftmargin=2em]
\item \textbf{Random:} uniform sampling over 110 bins.
\item \textbf{Global frequency:} always predict the most common \mz bin in the
  training set.
\item \textbf{Same-as-previous:} predict the same \mz bin as the preceding
  feature.
\item \textbf{Markov order-1 (\mz only):} \mz bin transition matrix estimated
  from training sequences.
\item \textbf{Markov order-2:} second-order \mz transition matrix.
\item \textbf{Frequency conditioned on current bin:} most frequent next \mz bin
  given the current \mz bin.
\item \textbf{Linear \rt extrapolation:} fit a linear model to the \mz values
  of the previous 5 or 10 features and extrapolate.
\item \textbf{Joint (\rt, \mz) Markov:} first-order transition matrix over
  (\rt bin, \mz bin) pairs, the strongest baseline at 56.8\% top-1 accuracy.
\end{enumerate}

\subsection{Neural Architectures}

Both neural models shared a common input embedding layer
(MultiFieldEmbedding) that sums five separate learned embeddings---one for each
input field (\mz bin, mass defect bin, \rt gap, polarity, intensity rank)---each
projecting to a 64-dimensional space. The summed embedding produces a single
64-dimensional vector per token.

\textbf{LSTM.}
A two-layer long short-term memory network with hidden dimension 128 and
dropout 0.1 (256{,}824 parameters). Tokens were processed sequentially; the
hidden state served as a compressed memory of all prior context in the elution
sequence. A linear projection from the 128-dimensional hidden state to 110
classes produced the output logits.

\textbf{Transformer.}
A two-layer Transformer encoder with a causal attention mask, 4 attention heads,
feed-forward dimension 256, and dropout 0.1 (121{,}784 parameters).\cite{vaswani2017attention}
Learned positional embeddings provided sequence-order information. The causal mask
prevented attention to future tokens, enforcing the autoregressive constraint.
The representation at the final context position was projected to 110 classes.

\textbf{Training.}
Both models were trained with the AdamW optimizer (learning rate $10^{-3}$,
weight decay $10^{-2}$), a cosine annealing learning rate scheduler, batch size
32, and cross-entropy loss over 110 \mz bin classes. Training ran for the full
100 epochs (early stopping with patience 10 was enabled but never triggered);
validation loss trended steadily downward for both architectures. The context window was 64 tokens, corresponding to approximately
192~seconds of elution at typical feature density. All training was performed on
Google Colab using an NVIDIA Tesla T4 GPU. The LSTM required approximately 5
hours and the Transformer approximately 7 hours to train.

\subsection{Ablation Study}

To quantify the contribution of each input feature and of sequence context
itself, we conducted an ablation study using embedding zeroing: for each
condition, the learned embedding of a specific input field was replaced with a
zero vector while all other model parameters remained identical. This approach
maintains the same parameter count across conditions, isolating the information
contributed by each field. Ten conditions were evaluated:

\begin{enumerate}[leftmargin=2em]
\item Full model (reference).
\item Five single-feature removals: $-$\mz bin, $-$mass defect, $-$\rt gap,
  $-$polarity, $-$intensity rank.
\item No-sequence (context window = 1 token), which removes autoregressive
  context entirely and reduces the model to a single-token classifier.
\item Three data-efficiency conditions: 25\%, 50\%, and 75\% of training data,
  to characterize the learning curve.
\end{enumerate}

Each condition was trained for 100 epochs with identical hyperparameters.

\subsection{QC Warm-Up Experiment}

This experiment tested whether LSTM hidden states could be ``primed'' by
processing QC injection sequences before evaluating analytical samples,
analogous to in-context learning in language models. The LSTM was selected for
this experiment because its recurrent hidden state provides a natural mechanism
for state conditioning; the Transformer's fixed-length context window makes
analogous conditioning less straightforward.

Four conditions were compared: (1)~cold start, with no prior conditioning;
(2)~prime-only, in which the LSTM processed a QC injection sequence and its
final hidden state was used as the initial state for the first prediction
position only; (3)~carry-hidden, in which the hidden state was propagated
through the entire analytical sample; and (4)~both prime and carry-hidden
combined. A dose-response analysis varied the number of conditioning QC
injections ($N = 0, 2, 4, 6, 8, 10$). Three control conditions were included:
analytical sample warm-up (conditioning on an analytical rather than QC sample),
cross-cohort QC warm-up (conditioning on QC injections from a different cohort),
and shuffled QC warm-up (conditioning on feature-order-randomized QC sequences).

\subsection{Dark Lipidome Annotation}

To assess whether model predictions could constrain database search for
unannotated features, we applied dual mass and retention time filtering against
two reference databases. For each of the 13{,}266 unannotated features, we
identified candidate annotations from the LIPID MAPS Structure Database (CC0
subset) and the MS-DIAL LipidBlast in-silico library that fell within
$\pm$10~ppm of the observed \mz and within $\pm$0.5~min of the observed
retention time. Of the 4{,}700 unannotated features whose \mz values fell
within the range covered by at least one reference entry, we reported the
number of unique putative annotations recovered. This analysis yields
Schymanski confidence level~3 (tentative candidate) annotations,\cite{schymanski2014confidence}
which require orthogonal confirmation.

\subsection{Cross-Platform Validation}

Three validation experiments assessed generalization beyond the training
cohorts, each isolating a different source of variation.

\textbf{Same chromatography, different instrument.}
ST000983 was acquired on an Agilent 6530 QTOF using the same CSH C18
chromatographic method as the training cohorts. We matched 242 lipids between
ST000983 and the training cohorts by normalized annotated lipid name and computed the Pearson
correlation and mean absolute error of retention times. This quantifies the
degree to which elution order is conserved across mass spectrometer platforms
when chromatographic conditions are held constant.

\textbf{Different chromatography.}
ST003514 was acquired on an Agilent 6545 QTOF with a different C18 column
chemistry. Data were tokenized using the same pipeline as training data
and evaluated with the trained LSTM and Transformer models without retraining or
fine-tuning. This tests the expected failure mode: models trained on one
chromatographic method should not generalize to a different method, because the
underlying elution physics differs.

\textbf{Raw data reprocessed.}
ST000990 was acquired on a SCIEX TripleTOF 6600 (same instrument family as
training data) with CSH C18 chromatography. We downloaded 153 raw .wiff files
(126 plasma, 15 QC, 10 technical replicates, 2 NIST SRM~1950 reference
samples), reprocessed them with MS-DIAL using an equivalent processing workflow, and
applied the trained models to the resulting feature table. This tests
end-to-end generalization from raw data through independent processing to
prediction.

\subsection{Transfer-Learning Recovery}

To test whether the cross-polarity failure on ST000990 reflects a recoverable
adaptation gap rather than a fundamental limitation, we measured how much
performance a small amount of calibration data from the new acquisition mode
restores. The 15 positive-mode QC injections of ST000990 served as a calibration
pool and its 126 analytical samples as a strictly held-out evaluation set
($n = 319{,}339$ predictions); the two sets are sample-disjoint. Before any
adaptation, we reproduced the zero-shot baseline (top-1 $= 0.026$) and confirmed
that the unmodified pretrained model scored at the 2.6\% floor on the held-out
analytical set, verifying that no calibration information leaks into evaluation.

We compared three adaptation strategies of increasing capacity as a function of
the number of calibration QC injections $N \in \{1, 2, 5, 10, 15\}$:
\begin{enumerate}[leftmargin=2em]
\item \textbf{Markov calibration} (no training): a first-order next-\mz-bin
  transition prior estimated from the $N$ QC injections, linearly blended with
  the pretrained model's output distribution,
  $p = \alpha\,p_{\mathrm{LSTM}} + (1 - \alpha)\,p_{\mathrm{Markov}}$ with
  $\alpha \in \{0, 0.25, 0.5, 0.75, 1\}$.
\item \textbf{Head-only fine-tuning}: the multi-field embedding and LSTM weights
  are frozen and only the final linear classification layer is retrained on the
  $N$ QC injections.
\item \textbf{Full fine-tuning}: all parameters are unfrozen and fine-tuned
  (Adam, learning rate $3 \times 10^{-4}$, batch size 32, up to 30 epochs),
  selecting the best epoch by accuracy on a held-out QC validation split
  (a separate QC injection for $N < 15$; an internal QC window for $N = 15$).
\end{enumerate}
All experiments used fixed random seeds (42) and were evaluated with the same
next-\mz-bin top-$k$ metrics as the within-method test set. The complete
pipeline---zero-shot reproduction, sample-disjointness and leakage checks, and
all three recovery methods---is provided as an executable, GPU-ready notebook in
the project repository.

\subsection{Evaluation Metrics}

The primary metric was top-1 accuracy on next \mz bin prediction (110 classes),
measuring the fraction of predictions where the highest-probability bin matched
the observed \mz bin. Secondary metrics included top-3, top-5, and top-10
accuracy (fraction of cases where the correct bin appeared among the $k$
highest-probability predictions). Error magnitude was quantified as the mean
absolute error in Da between the centers of the predicted and observed \mz bins.
Top-5 accuracy was highlighted as the most acquisition-relevant metric: in a
predictive acquisition scenario, the mass spectrometer could pre-configure
MS/MS parameters for five candidate \mz values, and top-5 accuracy estimates
the fraction of features for which the correct target would be among those
candidates.


\section{Results}

\subsection{Dataset Characterization}

Consensus feature alignment across the four clinical cohorts (REDHART~1, Cardiac Arrest, GVHD, PCOS) yielded 15{,}242 features from 342 human plasma samples acquired on the same LC-MS platform (SCIEX TripleTOF 6600+, Waters CSH C18 column). Of these, 1{,}976 (13\%) received putative lipid annotations via MS-DIAL with the LipidBlast library, spanning 12 lipid classes: acylcarnitine, lysophosphatidylethanolamine (LPE), lysophosphatidylcholine (LPC), fatty acid (FA), phosphatidylinositol (PI), phosphatidylcholine (PC), phosphatidylethanolamine (PE), sphingomyelin (SM), diacylglycerol (DG), ceramide (Cer), triacylglycerol (TG), and cholesteryl ester (CE). The remaining 87\% of features constitute the ``dark lipidome''---detected ions with no structural assignment. Annotation rates were consistent across cohorts (10.7--15.5\%), indicating that the annotation gap is a property of the analytical workflow rather than any individual cohort.

Cross-cohort retention time (\rt) reproducibility was exceptionally high. All six pairwise cohort comparisons yielded Pearson correlations of $r > 0.999$, with a mean absolute \rt error of 4.5~s across matched features. The GVHD--REDHART~1 pair showed the tightest agreement ($r = 1.000$, MAE = 0.6~s), consistent with acquisition in consecutive analytical batches. Cardiac Arrest pairs exhibited slightly higher MAE values (7.4--8.2~s), attributable to a 4-month temporal gap between acquisition campaigns. These results confirm that the CSH C18 chromatographic method produces a highly reproducible elution order across independent biological cohorts and acquisition dates.

\subsection{Elution Order Analysis}

To understand the physicochemical basis of elution order in reversed-phase lipidomics, we examined the relationship between \rt and molecular descriptors available for annotated features. Equivalent carbon number (ECN), a measure of hydrophobicity that accounts for both acyl chain length and unsaturation, correlated moderately with \rt ($r = 0.523$). However, lipid head group class introduced systematic \rt offsets from the ECN-predicted elution position, ranging from $-$169~s for LPC to $+$389~s for CE. Head group class alone explained $R^2 = 0.930$ of \rt variance, while ECN alone explained only $R^2 = 0.274$. A combined model incorporating ECN, head group class, and mass defect achieved $R^2 = 0.989$ (MAE = 14~s), confirming that elution order in reversed-phase lipidomics is governed by the interplay of hydrophobicity and polar head group identity.

Critically, lipid classes overlap extensively in \rt space. PC, PE, SM, DG, and Cer all co-elute between 5 and 6.5~min, and lipids sharing the same ECN but belonging to different classes can differ by up to 9.6~min in \rt. This mixing implies that predicting the next eluting species requires information about both hydrophobicity and head group identity---neither alone is sufficient.

For the 87\% of features lacking structural annotation, direct head group and ECN information is unavailable. However, mass defect (the fractional component of \mz) serves as a proxy: it encodes hydrogen content, which reflects chain saturation and thus hydrophobicity. Among annotated features, mass defect was the strongest univariate \rt predictor ($r = 0.899$). Even across all features including unannotated ions, mass defect retained moderate predictive power ($r = 0.377$). These observations motivated our tokenization design, which encodes \mz bin, mass defect, \rt gap, polarity, and intensity---features available for every detected ion regardless of annotation status.

\begin{figure}[!tb]
\centering
\includegraphics[width=\textwidth]{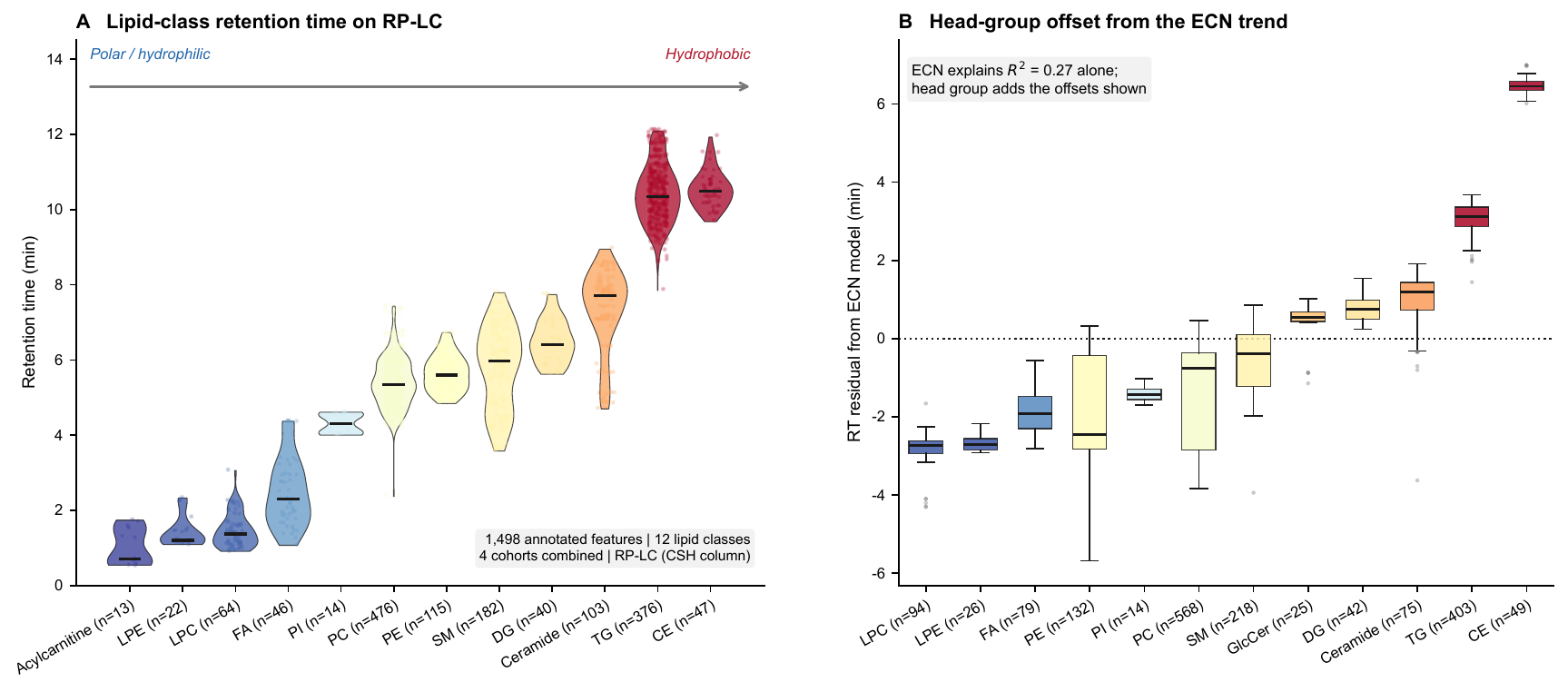}
\caption{\textbf{Physicochemical basis of elution order in reversed-phase lipidomics.}
\textbf{(A)}~Retention time distributions for 12 lipid classes across all four training
cohorts ($n = 1{,}498$ annotated features). Classes are ordered from most polar
(acylcarnitine, left) to most hydrophobic (cholesteryl ester, right). Horizontal bar
indicates median; individual points are jittered within each violin. Note the extensive
overlap between PC, PE, and SM in the 4--7~min region.
\textbf{(B)}~Systematic head group contribution to retention time. Boxplots show
residuals from a global linear model of \rt~vs.~equivalent carbon number (ECN), grouped by
lipid class ($n = 1{,}752$). Polar head groups (LPC, FA, LPE) elute earlier than
ECN predicts; nonpolar structures (TG, CE) elute later. The span from LPC ($-$169~s) to
CE ($+$389~s) confirms that head group identity introduces systematic \rt offsets of up
to 9.6~min at the same ECN.}
\label{fig:elution_physics}
\end{figure}

\subsection{Baseline Performance}

We evaluated eight baseline strategies for next-\mz-bin prediction, a 110-class classification task where each class corresponds to a 10~Da \mz bin; the five most informative are reported in Table~\ref{tab:baselines}. Results are reported on the held-out test set ($n = 186{,}052$ tokens).

\begin{table}[htbp]
\centering
\caption{Baseline performance on next-\mz-bin prediction (110 classes, test set).}
\label{tab:baselines}
\begin{tabular}{lccccr}
\toprule
Baseline & Top-1 & Top-3 & Top-5 & Top-10 & MAE (Da) \\
\midrule
Random                  & 1.5\%  & ---    & ---    & ---    & 280 \\
Global frequency        & 3.4\%  & ---    & ---    & ---    & 203 \\
Same-as-previous \mz    & 23.1\% & ---    & ---    & ---    & 119 \\
Markov order-1 (\mz)    & 25.1\% & 47.9\% & 58.3\% & 72.3\% & 115 \\
Joint (\rt, \mz) Markov & \textbf{56.8\%} & ---    & ---    & ---    & 68 \\
\bottomrule
\end{tabular}
\end{table}

The joint (\rt, \mz) Markov baseline was the strongest, achieving 56.8\% top-1 accuracy by conditioning on both the current \rt position and the most recent \mz bin. This doubled the accuracy of the \mz-only Markov model (25.1\%), demonstrating that chromatographic position carries substantial information about which \mz bins are likely to appear next. Nevertheless, an oracle model given the true \rt of the next feature but no sequence history achieved only 14.1\% top-1 accuracy, confirming that \rt position alone is insufficient---sequential context is essential.

The same-as-previous baseline (23.1\%) established that adjacent features share the same \mz bin less than a quarter of the time, confirming that next-\mz-bin prediction is a genuinely challenging task with high class diversity in the local elution neighborhood.

\subsection{Autoregressive Model Performance}

Both neural architectures substantially outperformed all baselines (Table~\ref{tab:models}). The LSTM (256{,}824 parameters) achieved 98.38\% top-1 accuracy and 99.99\% top-5 accuracy on the test set, with a mean absolute error of 3.6~Da. The Transformer (121{,}784 parameters) achieved 98.05\% top-1 and 99.88\% top-5 accuracy with an MAE of 4~Da.

\begin{table}[htbp]
\centering
\caption{Autoregressive model performance on next-\mz-bin prediction (test set, $n = 182{,}902$).}
\label{tab:models}
\begin{tabular}{lrccc}
\toprule
Model & Parameters & Top-1 & Top-5 & MAE (Da) \\
\midrule
Joint Markov (best baseline)  & --- & 56.8\%  & ---     & 68 \\
Transformer                   & 121{,}784 & 98.05\% & 99.88\% & 4 \\
LSTM                          & 256{,}824 & \textbf{98.38\%} & \textbf{99.99\%} & \textbf{3.6} \\
\bottomrule
\end{tabular}
\end{table}

Both models achieved more than 41 percentage points of improvement over the strongest baseline. When predictions were incorrect, the predicted \mz bin was typically adjacent to the true bin: the 3--4~Da MAE at a 10~Da bin width indicates that most errors are off-by-one bin misclassifications rather than gross prediction failures. The LSTM held a modest advantage over the Transformer (+0.33 percentage points top-1), suggesting that the sequential inductive bias of recurrent processing aligns well with the inherently sequential nature of chromatographic elution.

Training dynamics were stable for both architectures. The LSTM achieved 99.9\% top-5 accuracy by epoch~6, and validation loss decreased steadily over all 100~epochs without triggering early stopping. The LSTM required 5.1~hours to train on a single T4 GPU. The Transformer converged somewhat faster in early epochs but plateaued at a slightly lower final accuracy, despite using roughly half the parameters of the LSTM. Both architectures demonstrated that chromatographic elution sequences contain highly learnable structure.

\begin{figure}[!tb]
\centering
\includegraphics[width=\textwidth]{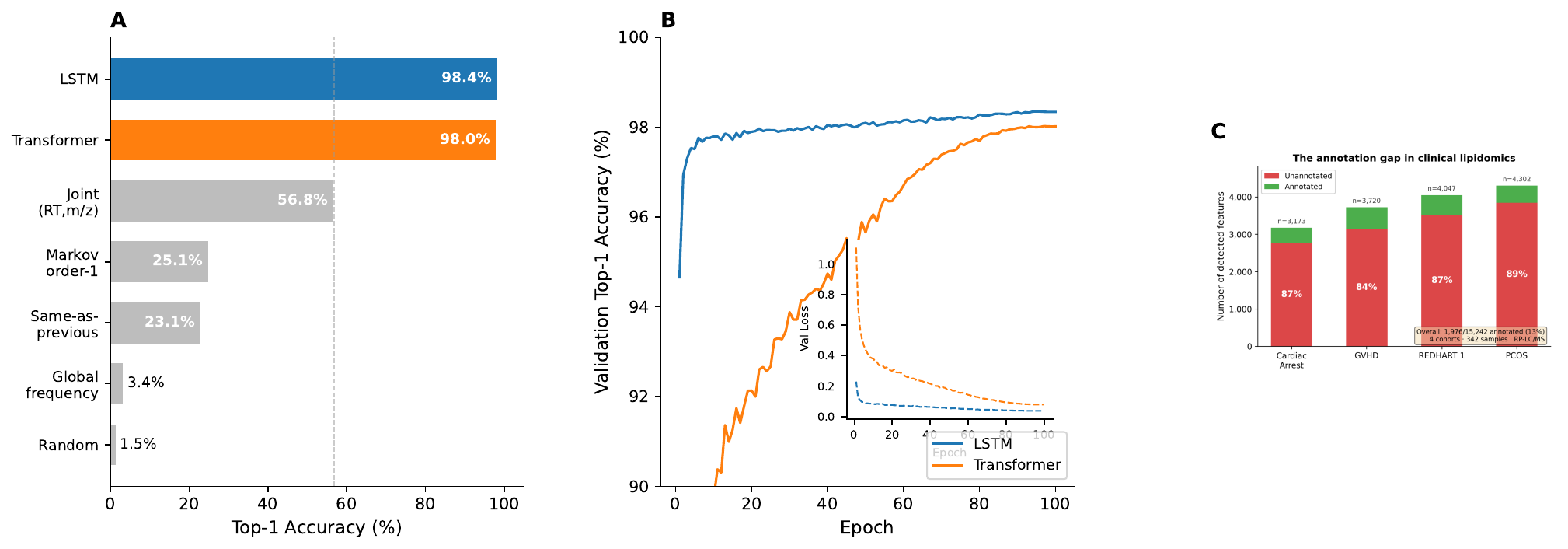}
\caption{\textbf{Model performance and dataset context.}
\textbf{(A)}~Top-1 accuracy on next-\mz-bin prediction (110 classes, test set;
neural models $n = 182{,}902$, baselines $n = 186{,}052$). The LSTM (98.38\%) and Transformer (98.05\%) far exceed all
baselines, with the joint (\rt, \mz) Markov model (56.8\%) as the strongest
non-neural comparator. Error bars indicate 95\% bootstrap confidence intervals.
\textbf{(B)}~Training dynamics. Validation top-1 accuracy (solid) and validation
loss (dashed) over 100 epochs for both architectures. The LSTM converges to a
slightly higher asymptote; both models show an overall downward validation-loss
trend with no overfitting.
\textbf{(C)}~Annotation gap across the four training cohorts. Each bar shows the
number of detected features, colored by annotation status. Overall, only 13\%
of 15{,}242 consensus features received structural annotations, leaving 87\% as
the ``dark lipidome.''  Annotation rates are consistent across cohorts
(10.7--15.5\%), indicating a workflow-level limitation.}
\label{fig:model_performance}
\end{figure}

\subsection{Feature Ablation}

To quantify the contribution of each input feature to prediction accuracy, we conducted a systematic ablation study using an embedding-zeroing approach. This strategy replaces the embedding vector for the ablated feature with zeros while keeping all model parameters identical, ensuring that differences in accuracy reflect information content rather than architectural changes. Ten conditions were tested: the full model, five single-feature removals (\mz bin, mass defect, \rt gap, polarity, intensity rank), a no-sequence condition (context window = 1 token), and three data-efficiency conditions (25\%, 50\%, and 75\% of training data).

\begin{figure}[!tb]
\centering
\includegraphics[width=\textwidth]{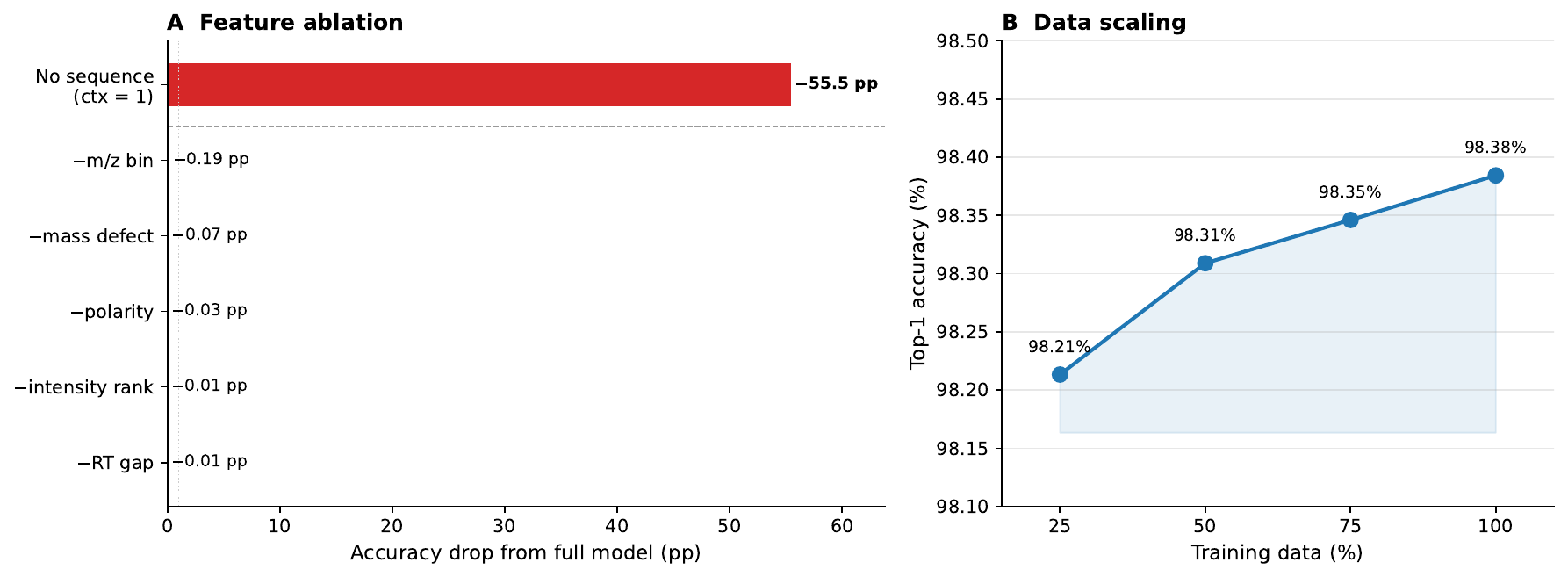}
\caption{\textbf{Feature ablation and data efficiency.}
\textbf{(A)}~Accuracy drop from the full model when each input feature is removed
via embedding zeroing (blue bars) or when autoregressive context is eliminated
(red bar, context window = 1 token). Removing sequence context causes a
55.5~percentage-point drop; no individual feature contributes more than 0.19~pp.
\textbf{(B)}~Data efficiency curve. Top-1 accuracy as a function of training set
size (25\%, 50\%, 75\%, 100\%). Performance plateaus rapidly, with 25\% of data
sufficient to achieve 98.2\% accuracy.}
\label{fig:ablation}
\end{figure}

Sequence context was overwhelmingly the dominant source of predictive information (Figure~\ref{fig:ablation}A, Table~\ref{tab:ablation}). Removing autoregressive context entirely (context window = 1 token) reduced top-1 accuracy from 98.38\% to 42.89\%---a drop of 55.5 percentage points---and inflated the mean absolute error from 3.6 to 91.2~Da. Without the preceding elution history, the model has essentially no ability to predict what comes next, despite retaining all five per-token input features.

\begin{table}[htbp]
\centering
\caption{Ablation study results (LSTM, test set, $n = 182{,}902$). Embedding zeroing removes one input field at a time; no-sequence reduces context to 1 token.}
\label{tab:ablation}
\begin{tabular}{lccc}
\toprule
Condition & Top-1 (\%) & $\Delta$ Top-1 (pp) & MAE (Da) \\
\midrule
Full model (reference)  & 98.38 & ---     & 3.6 \\
$-$\mz bin              & 98.20 & $-$0.19 & 3.8 \\
$-$Mass defect          & 98.31 & $-$0.07 & 3.6 \\
$-$RT gap               & 98.37 & $-$0.01 & 3.6 \\
$-$Polarity             & 98.36 & $-$0.03 & 3.6 \\
$-$Intensity rank       & 98.37 & $-$0.01 & 3.6 \\
No sequence (ctx = 1)   & 42.89 & $-$55.49 & 91.2 \\
\midrule
25\% training data      & 98.21 & $-$0.17 & 3.9 \\
50\% training data      & 98.31 & $-$0.07 & 3.8 \\
75\% training data      & 98.35 & $-$0.04 & 3.7 \\
\bottomrule
\end{tabular}
\end{table}

By contrast, removing any single input feature produced minimal accuracy loss. The \mz bin was the most informative individual feature ($-$0.19 percentage points when removed), followed by mass defect ($-$0.07~pp). Polarity, intensity rank, and retention-time gap each contributed $\leq$0.03~pp. The asymmetry is striking: the autoregressive context accounts for more than 99\% of the accuracy gain over a single-token classifier, while all five per-token features combined contribute about 0.3~pp. This indicates that the model's predictive power derives almost entirely from the sequential pattern of preceding features rather than from the observable properties of any individual token.

The data efficiency curve (Figure~\ref{fig:ablation}B) showed that the model is remarkably sample-efficient. Training on 25\% of the data (approximately 60 samples) still achieved 98.21\% top-1 accuracy---only 0.17~pp below the full model. The learning curve exhibited diminishing returns, with the largest gain between 25\% and 50\% (0.10~pp) and a marginal gain from 75\% to 100\% (0.04~pp). These results suggest that elution sequence patterns are sufficiently regular that a modest number of samples captures the essential structure.

\subsection{QC Warm-Up: A Null Result}

Quality control (QC) injections are routinely used at the beginning of LC-MS analytical sequences to condition the chromatographic column and stabilize the instrument response\cite{broadhurst2018qc}. We hypothesized that feeding QC injection sequences through the LSTM before evaluating test samples might condition the hidden state with method-specific context, analogous to in-context learning in large language models. We tested four conditions: cold start (no QC priming), prime-only (process QC sequences but reset hidden state before test), carry-hidden (retain QC-conditioned hidden state into test inference), and both (prime and carry).

\begin{figure}[!tb]
\centering
\includegraphics[width=\textwidth]{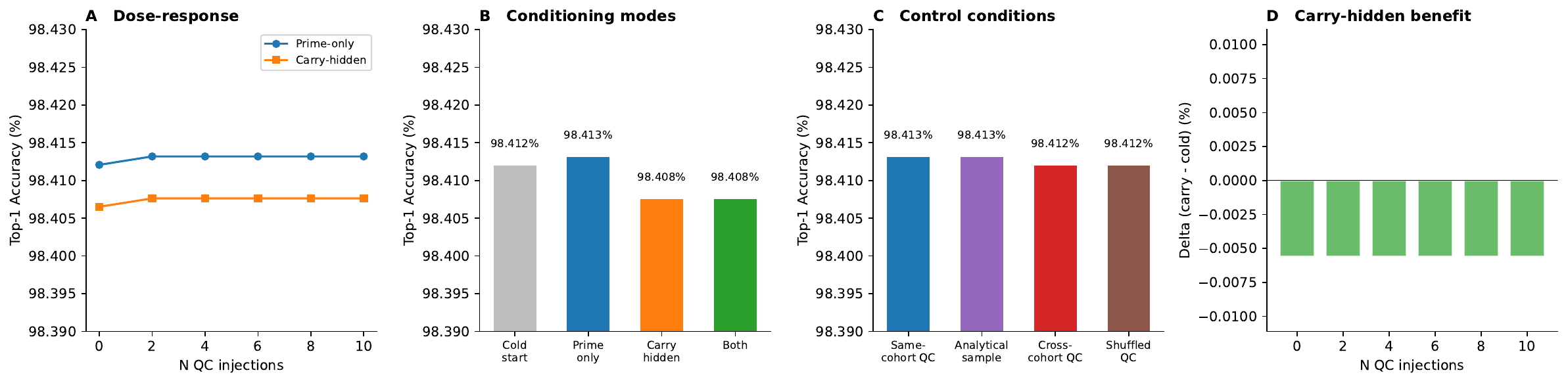}
\caption{\textbf{QC warm-up experiment: a null result.}
\textbf{(A)}~Dose--response analysis. Top-1 accuracy as a function of the number
of QC injection sequences used to condition the LSTM hidden state before
evaluation ($N = 0$--$10$). The curve is flat, with no benefit from additional
QC conditioning (maximum delta = 0.001\%).
\textbf{(B)}~Comparison of four conditioning modes: cold start, prime-only (use
QC-conditioned hidden state for first prediction only), carry-hidden (propagate
hidden state through entire sample), and both. All conditions produce
statistically indistinguishable accuracy.
\textbf{(C)}~Control conditions: conditioning on same-cohort QC, an analytical
sample, cross-cohort QC, or order-shuffled QC injections all yield
indistinguishable accuracy (${\sim}98.41\%$), confirming the null result is
robust to the type of priming data.
\textbf{(D)}~Carry-hidden benefit: the accuracy delta (carry-hidden minus cold
start) as a function of the number of QC injections fluctuates around zero
(within $\pm0.01\%$), confirming the absence of a warm-up benefit.}
\label{fig:qc_warmup}
\end{figure}

The result was unambiguously null. The maximum accuracy difference between any conditioning mode and cold start was 0.001\% (Figure~\ref{fig:qc_warmup}). A dose--response experiment across 0--10~QC injections produced a flat curve: additional QC sequences provided no incremental benefit. Cross-cohort QC conditioning (using QC sequences from a different cohort than the test samples) likewise produced no improvement. Even shuffled QC sequences---which destroy temporal structure---performed identically to ordered QC sequences.

This null result carries a clear interpretation: the LSTM captures the elution logic of the chromatographic method entirely from training data, without requiring in-context adaptation at inference time. The model has internalized the relationship between token features and elution position during training, and the hidden state converges to a useful representation within the first few tokens of any new sequence regardless of prior conditioning.

The practical implication is significant for deployment. No special QC conditioning protocol is required before applying the model to new samples. Cold start inference is sufficient, simplifying integration into automated acquisition workflows.

\subsection{Cross-Platform Validation}

To assess the generalizability of trained models, we evaluated performance under three transfer conditions that systematically varied the degree of analytical method similarity (Table~\ref{tab:crossplatform}).

\begin{table}[htbp]
\centering
\caption{Cross-platform validation results. \rt correlation and model prediction accuracy under varying degrees of method transfer.}
\label{tab:crossplatform}
\begin{tabular}{p{3.2cm}p{3cm}p{2.5cm}cc}
\toprule
Condition & Dataset & Chromatography & \rt~$r$ & Top-1 \\
\midrule
Within-training & 4 cohorts & CSH C18 (same) & $>$0.999 & 98.4\% \\
Same method, different MS & ST000983 (Agilent 6530) & CSH C18 (same) & 0.9993 & --- \\
Different method & ST003514 (Agilent 6545) & Different C18 & --- & 5.1\% \\
Same instrument, single polarity & ST000990 (SCIEX 6600) & CSH C18 (same) & --- & 2.6\% \\
\bottomrule
\end{tabular}
\end{table}

\subsubsection{Same chromatography, different mass spectrometer (ST000983).}
\begin{figure}[!tb]
\centering
\includegraphics[width=\textwidth]{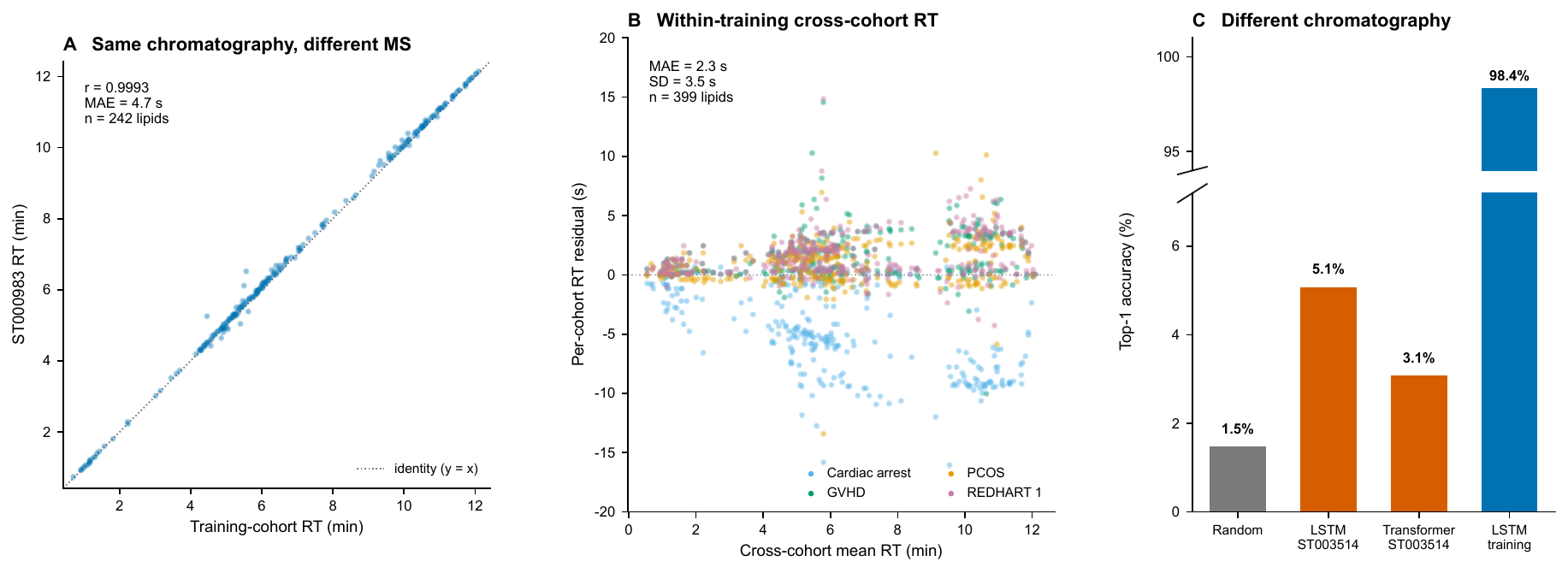}
\caption{\textbf{Cross-platform validation: models are chromatographic-method-specific.}
\textbf{(A)}~Retention time correlation between the training cohorts and ST000983
(Agilent 6530 QTOF, same CSH C18 chromatographic method). Each point represents
one of 242 matched lipids ($r = 0.9993$, MAE = 4.7~s). The near-perfect
correlation demonstrates that elution order transfers across mass spectrometer
platforms when the chromatographic method is held constant.
\textbf{(B)}~Within-training cross-cohort \rt consistency: for each of 399 lipids
detected in at least two training cohorts, the per-cohort median \rt residual
relative to that lipid's cross-cohort mean, coloured by cohort (MAE = 2.3~s,
SD = 3.5~s). Residuals cluster tightly around zero across the full elution
window, demonstrating a stable elution order across independent biological
cohorts (all six pairwise cohort correlations $r > 0.999$; tightest GVHD vs.\
REDHART~1 at $r = 1.000$; see text).
\textbf{(C)}~Model accuracy on ST003514 (Agilent 6545 QTOF, different C18 column
chemistry). The LSTM achieves 5.1\% and the Transformer 3.1\% top-1 accuracy,
near-random performance on a 110-class task, confirming that models do not
transfer across chromatographic methods. The ST000990 polarity-mode failure
(2.6\% top-1) is described in the text.}
\label{fig:cross_platform}
\end{figure}

ST000983\cite{cajka2017validating} was acquired on an Agilent 6530 QTOF using the same Waters CSH C18 column and mobile phase system as the training cohorts. We identified 242 lipids present in both ST000983 and the training cohorts by matching normalized annotated lipid names. Retention times correlated at $r = 0.9993$ with a mean absolute error of 4.7~s (Figure~\ref{fig:cross_platform})---comparable to the within-training-set cross-cohort MAE of 4.5~s (the mean of the six pairwise cohort comparisons). This result demonstrates that elution order is defined by the chromatographic method (\ie the stationary phase, mobile phase composition, and gradient program), not by the mass spectrometer. Models trained on data from one instrument can be applied to data from any instrument using the same chromatographic method without retraining.

\subsubsection{Different chromatography (ST003514).}
ST003514\cite{martinez2024srm1950} was acquired on an Agilent 6545 QTOF using a different C18 column with distinct stationary phase chemistry. We applied the trained LSTM and Transformer directly to tokenized ST003514 sequences without retraining. The LSTM achieved only 5.1\% top-1 accuracy and the Transformer only 3.1\%---near-random performance on a 110-class task (random baseline: 1.5\%). This catastrophic failure confirms that the models learn chromatographic-method-specific elution physics. A different column chemistry produces a fundamentally different elution order, and the learned sequential dependencies do not transfer.

\subsubsection{Same instrument, different polarity mode (ST000990).}
ST000990 was acquired on a SCIEX TripleTOF 6600 (the same instrument family as the training cohorts' 6600+) using the same CSH C18 chromatographic method.\cite{cajka2017validating} We reprocessed 153 raw .wiff files through MS-DIAL, yielding 3{,}604 consensus features (a mean of ${\sim}2{,}655$ detected per sample). Despite using the same instrument family and identical column chemistry, the LSTM achieved only 2.6\% top-1 accuracy (MAE $\approx$ 250~Da)---worse than the random baseline. The critical difference was polarity mode: ST000990 was acquired in positive-ESI only, whereas the training data were acquired in dual-polarity mode (alternating positive and negative scans). This created a fundamental mismatch in elution sequence structure. In dual-polarity runs, the model learns to predict polarity-mode transitions (positive $\to$ negative $\to$ positive) that account for a substantial fraction of the sequential pattern. In positive-only data, these transitions are absent, and the sequence statistics differ radically: median inter-feature \mz jumps were 3$\times$ larger in ST000990 (12 bins vs.\ 4 bins in training data).

This result reveals a polarity paradox: ST003514, acquired on a \textit{different} instrument with \textit{different} column chemistry but in dual-polarity mode, achieved 5.1\% top-1 accuracy---nearly double that of ST000990, which used the \textit{same} instrument family and \textit{same} column. The model is not merely method-specific; it is polarity-mode-specific. The sequential structure of the elution stream depends on which ion modes are interleaved during acquisition, not solely on the chromatographic separation.

\subsubsection{Synthesis.}
These results establish a clear hierarchy of transferability. Models generalize perfectly across mass spectrometers using the same chromatographic method and polarity mode ($r = 0.9993$, MAE = 4.7~s). They fail when the column chemistry changes (ST003514: 5.1\% top-1), and they fail even more severely when the polarity acquisition mode changes, even on the same instrument and column (ST000990: 2.6\% top-1). The model learns the elution physics of a specific chromatographic method \textit{as expressed through a specific acquisition mode}. For deployment, a single trained model can serve all instruments sharing the same method and polarity mode, but changes to either require retraining or transfer learning.

\subsection{Transfer Learning Recovers Cross-Polarity Performance}

\begin{figure}[!tb]
\centering
\includegraphics[width=0.85\textwidth]{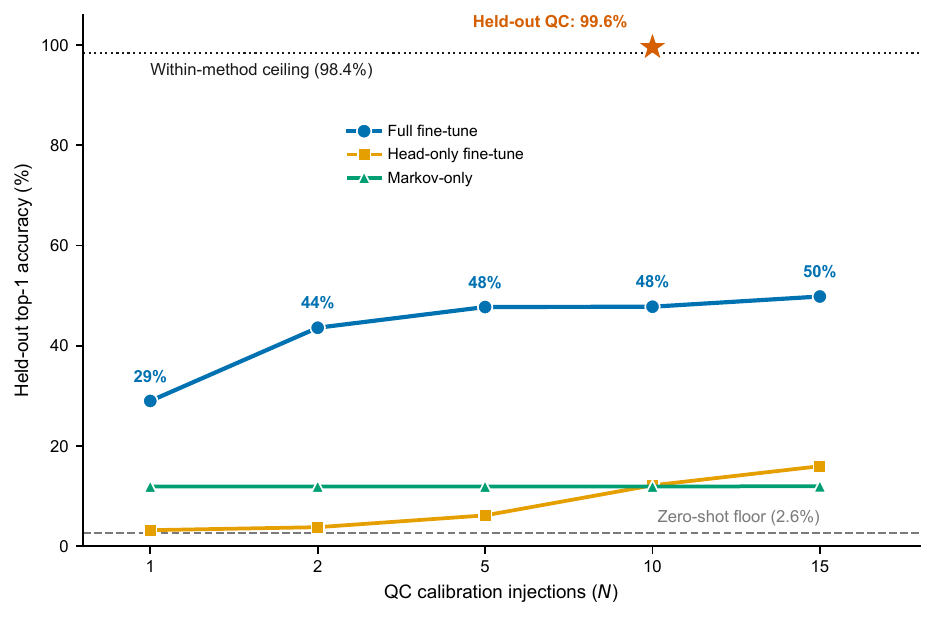}
\caption{\textbf{Transfer learning recovers cross-polarity performance.}
Held-out analytical top-1 accuracy on ST000990 (positive-ESI only) as a function
of the number of QC calibration injections $N$, for full fine-tuning (all
parameters), head-only fine-tuning (output layer only), and a calibration-only
Markov prior. Full fine-tuning recovers from the 2.6\% zero-shot floor (dashed)
toward $\sim$50\%---and to 99.6\% on a held-out QC injection (vermillion
star)---while head-only tuning and the Markov prior plateau below 16\%,
localizing the cross-polarity deficit to the learned sequence representation. The
dotted line is the within-method ceiling (98.4\%) on matched dual-polarity data.}
\label{fig:recovery}
\end{figure}

The catastrophic zero-shot failure under positive-only acquisition proved to be a \emph{recoverable adaptation gap} rather than a fundamental limitation (Table~\ref{tab:recovery}, Figure~\ref{fig:recovery}). Full fine-tuning of the pretrained model on a small number of QC injections restored substantial accuracy on the strictly held-out analytical samples. With 15 QC injections, held-out analytical top-1 accuracy recovered from the 2.6\% zero-shot floor to \textbf{49.8\%} (top-5 = 67.9\%). On a genuinely held-out QC injection (fine-tuning on 10 of the 15 QC injections and evaluating on a withheld one), the model reached \textbf{99.6\%} top-1---it had fully relearned the positive-only elution grammar from minimal data. Recovery was rapid and data-efficient: as few as two QC injections lifted analytical top-1 to 43.6\%, and the curve plateaued near 48--50\% by five injections.

Critically, recovery required adapting the learned \emph{representation}, not merely recalibrating the output. Head-only fine-tuning (freezing the embedding and LSTM) plateaued at 16.0\% top-1, and a calibration-only Markov prior at 12.0\%, whereas full fine-tuning reached 49.8\%---localizing the cross-polarity deficit to the LSTM's internal sequence representation, which only full fine-tuning can re-tune. The residual gap between the near-perfect held-out-QC accuracy (99.6\%) and the held-out analytical accuracy (49.8\%) reflects the limited diversity of pooled QC injections relative to 126 biologically distinct analytical samples; calibration data spanning that variability would be expected to narrow it.

\begin{table}[htbp]
\centering
\caption{Transfer-learning recovery on held-out ST000990 analytical samples ($n = 319{,}339$ predictions). Zero-shot floor $= 2.6$\%; within-method ceiling $= 98.4$\%.}
\label{tab:recovery}
\begin{tabular}{lccccc}
\toprule
QC injections ($N$) & 1 & 2 & 5 & 10 & 15 \\
\midrule
Full fine-tune (top-1)      & 29.0\% & 43.6\% & 47.7\% & 47.8\% & \textbf{49.8\%} \\
Full fine-tune (top-5)      & 55.6\% & 64.7\% & 67.2\% & 66.9\% & 67.9\% \\
Head-only fine-tune (top-1) & 3.2\%  & 3.8\%  & 6.2\%  & 12.2\% & 16.0\% \\
Markov-only (top-1)         & 11.9\% & 11.9\% & 11.9\% & 11.9\% & 12.0\% \\
\bottomrule
\end{tabular}
\end{table}

\subsection{Dark Lipidome Annotation}

The 87\% of consensus features lacking structural annotation represent the ``dark lipidome''---ions that are reproducibly detected but not identified by conventional spectral matching\cite{daSilva2015darkMatter}. To assess whether these features correspond to known lipid species that were missed by the standard annotation pipeline, we applied a dual filtering strategy: mass tolerance of $\pm$10~ppm and \rt tolerance of $\pm$0.5~min against reference entries from LipidMaps (CC0-licensed subset) and the LipidBlast library used by MS-DIAL.

Of the 13{,}266 unannotated features, approximately 4{,}700 fell within the \mz range covered by the reference databases. From this subset, 168 unique putative lipid annotations were recovered. These represent features whose accurate mass and retention time match known lipid species but which were not annotated by the original MS-DIAL + LipidBlast pipeline---likely because MS2 fragmentation spectra were absent or scored below the annotation threshold. While these annotations remain putative (Schymanski confidence level~3\cite{schymanski2014confidence}), they demonstrate that combining mass accuracy with chromatographic retention time constraints can substantially narrow the search space for structural assignment of unannotated features.

\begin{figure}[!tb]
\centering
\includegraphics[width=\textwidth]{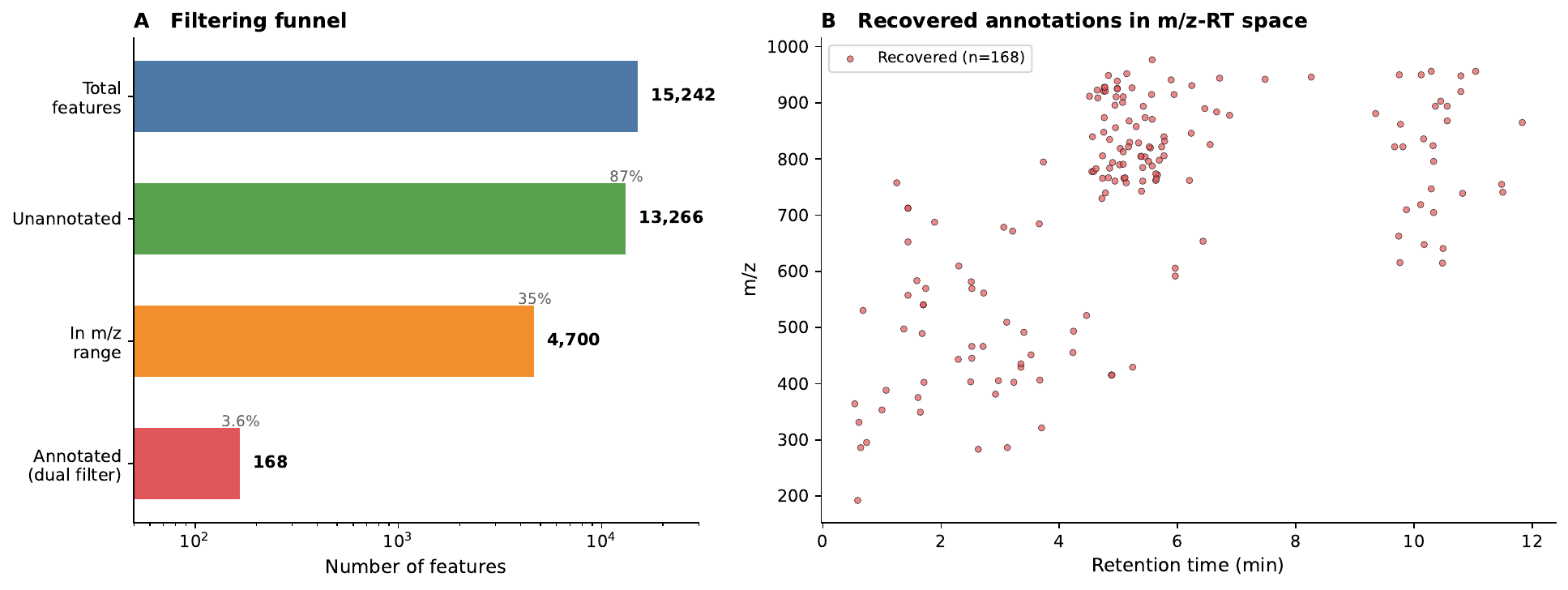}
\caption{\textbf{Dark lipidome annotation via dual mass and retention time filtering.}
\textbf{(A)}~Filtering funnel. Of 13{,}266 unannotated consensus features,
approximately 4{,}700 fell within the \mz range covered by the LipidMaps and
LipidBlast reference databases. Dual filtering ($\pm$10~ppm mass tolerance,
$\pm$0.5~min \rt tolerance) recovered 168 unique putative lipid annotations.
\textbf{(B)}~Distribution of recovered annotations (colored points) overlaid on
all unannotated features (gray) in \mz--\rt space. Recovered annotations span
the full chromatographic range, indicating that the approach is not biased
toward a specific elution region or mass range.}
\label{fig:dark_lipidome}
\end{figure}

\subsection{Sequence Position Analysis}

A potential concern with autoregressive models is that prediction accuracy might degrade at the beginning of a chromatographic run, where limited preceding context is available, or at the end, where rare late-eluting species may be underrepresented in training data. We examined top-1 accuracy as a function of sequence position across the entire chromatographic run.

\begin{figure}[!tb]
\centering
\includegraphics[width=0.85\textwidth]{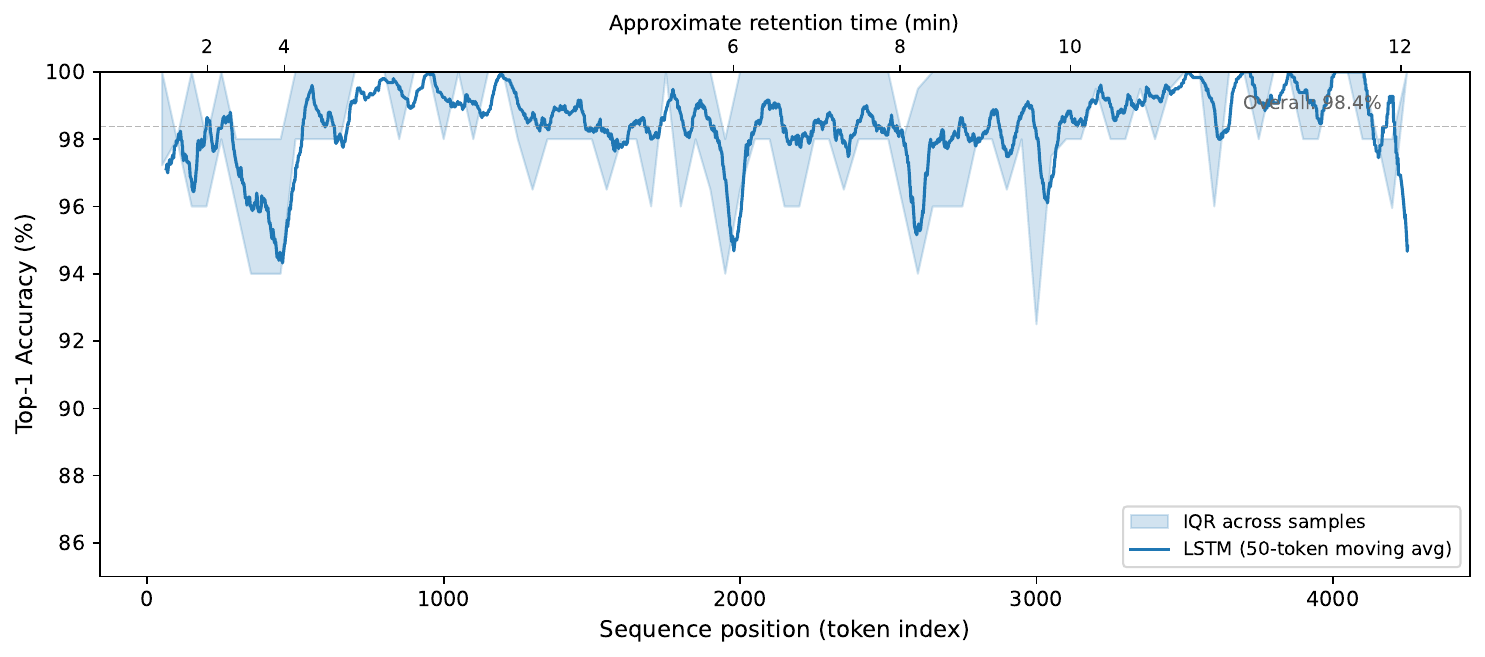}
\caption{\textbf{Position-dependent prediction accuracy.}
Top-1 accuracy on next-\mz-bin prediction as a function of sequence position
(token index within the chromatographic run) for the LSTM (blue line),
averaged across all test samples with a 50-token moving average.
Accuracy is approximately 98\% and uniform across the entire run, with no
systematic degradation at early positions (where the context window is not yet
fully populated) or late positions (where rare late-eluting species
predominate). The shaded region indicates the interquartile range across test
samples. This uniformity argues against position-specific memorization and
supports the interpretation that the model learns generalizable sequential
dependencies.}
\label{fig:position_accuracy}
\end{figure}

Accuracy was remarkably uniform, averaging approximately 98\% at all positions with no systematic degradation at early or late positions (Figure~\ref{fig:position_accuracy}). The model achieved high accuracy even for the first few dozen tokens, where the context window is not yet fully populated. This indicates that the model does not require a long ``burn-in'' period to build a useful internal representation of the elution state. Rather, the combination of absolute features (\mz bin, mass defect, polarity) and the first few sequential observations provides sufficient information for accurate prediction.

The absence of position-dependent accuracy variation also argues against a trivial memorization explanation. If the model had simply memorized position-specific \mz distributions, we would expect accuracy to track the entropy of the \mz distribution at each position---higher in regions of dense class overlap, lower in sparse regions. Instead, the uniform accuracy profile suggests that the model learns generalizable sequential dependencies that apply throughout the chromatographic run.

\section{Discussion}


The central finding of this work is that the chromatographic elution sequence
in untargeted LC-HRMS lipidomics is highly predictable from a compact set of
observable features, without any structural or spectral information. An LSTM
network achieved 98.4\% top-1 accuracy in predicting the next \mz bin, and a
Transformer achieved 98.0\%, both far exceeding the best classical
baseline (Joint Markov, 56.8\%). These accuracies were achieved using only five
per-token input features---\mz bin, mass defect, retention-time gap, ionization
polarity, and intensity rank---none of which require structural annotation. The
fact that near-perfect prediction is possible from such minimal observables
suggests that the elution sequence encodes its own predictive structure: the
``language'' of chromatographic elution is simple, regular, and learnable. Below
we discuss the implications of these results, their practical significance for
acquisition planning and annotation, and the boundaries of what has been
demonstrated.

The progression from simple baselines to deep sequence models reveals where
predictive information resides in the elution stream. An oracle that knows the
exact retention time of the next feature but nothing about \mz history achieves
only 14.1\% top-1 accuracy---knowing \textit{where} you are in the
chromatographic run is far from sufficient. The Joint Markov baseline, which
conditions on both the current retention-time bin and the most recent \mz bin,
reaches 56.8\%, demonstrating that short-range \mz context roughly doubles
prediction accuracy. The jump from 56.8\% to 98.4\% with the LSTM, which
processes 64 preceding tokens, reveals that extended sequential context contains
substantial predictive information inaccessible to pairwise statistics. This
finding parallels observations in natural language processing, where long-range
dependencies captured by recurrent and attention-based architectures account for
the qualitative superiority of neural language models over $n$-gram
approaches.\cite{vaswani2017attention} The chromatographic elution sequence, while
far simpler than natural language, is evidently rich enough in sequential
structure that modeling it as a token-prediction task is both natural and
effective. Importantly, the comparable performance of the LSTM (98.4\%) and
Transformer (98.0\%) suggests that the sequential dependencies in
chromatographic data are well captured by recurrent architectures and do not
require the full self-attention mechanism to exploit. Whether longer sequences
or more complex chromatographic separations would favor attention-based models
remains an open question.

The ablation study quantifies this dominance of sequential context. Removing
autoregressive history (context window = 1 token) drops accuracy by 55.5
percentage points, whereas removing any individual input feature costs at most
0.19~pp (\mz bin). The five per-token features---\mz bin, mass defect,
retention-time gap, polarity, and intensity rank---are nearly redundant once the
model has access to the preceding sequence. This is consistent with the
interpretation that the ``language'' of chromatographic elution is encoded in the
ordering of features, not in the properties of any single feature. The high data
efficiency (98.2\% accuracy with only 25\% of training data) further suggests
that the sequential patterns are regular and repetitive, requiring relatively few
examples to learn. These findings have practical implications: the model is
robust to noisy or missing feature-level annotations, and deployment on new
datasets requires fewer training samples than might be expected from the
complexity of the prediction task.

The physical basis for this predictability is hydrophobic partitioning. Mass
defect---the fractional component of \mz---correlates with retention time at
$r = 0.899$, making it the single strongest predictor among our input features.
This is not coincidental: mass defect encodes hydrogen content relative to
nominal mass, which in lipids directly reflects acyl-chain saturation and
length, \ie hydrophobicity.\cite{white2022ecn} Head-group class explains
$R^{2} = 0.930$ of retention-time variance, but this taxonomic variable is
largely a proxy for the underlying polarity differences between classes. When a
direct measure of hydrophobicity (MolLogP) is available for the annotated
subset, it captures 83.5\% of retention-time variance and renders class nearly
redundant. Critically, 87\% of features in our dataset lack structural
annotation, so explicit hydrophobicity measures are unavailable. The model
compensates by extracting hydrophobicity information implicitly from mass defect,
\mz, and the surrounding elution context. In effect, the model learns the
physics of reversed-phase partitioning\cite{vankova2022retention} from
observables alone, without being told what the analytes are.

The cross-platform validation experiments provide the clearest evidence that these
models learn chromatographic physics rather than instrument-specific artifacts.
When the chromatographic method is held constant (same CSH C18 column and mobile
phase) but the mass spectrometer changes from a SCIEX TripleTOF 6600+ to an
Agilent 6530 QTOF, the retention-time correlation between training and
validation data is $r = 0.9993$ with a mean absolute error of 4.7~seconds---the
model transfers almost perfectly. In contrast, when the column chemistry changes
(ST003514, a different C18 stationary phase with different mobile-phase
modifiers), top-1 accuracy collapses to 5.1\%, little better than chance. This
dichotomy is the expected behavior: reversed-phase selectivity is determined by
the stationary-phase chemistry and mobile-phase
composition,\cite{bach2018retention} not by the detector downstream. The same
column and gradient produce the same elution order regardless of which mass
spectrometer records the ions. The practical implication is that one trained model
serves all instruments running a given chromatographic method. Laboratories
sharing the same column, gradient, and mobile phase can share a single predictive
model, while a change in chromatographic method requires retraining. This is
analogous to how a language model trained on English transfers across keyboards
but not across languages.

The ST000990 validation experiment revealed an additional dimension of method
specificity: polarity acquisition mode. Despite sharing the same instrument
family (SCIEX TripleTOF 6600) and the same CSH C18 column as the training data,
ST000990 yielded only 2.6\% top-1 accuracy---worse even than the random
baseline. The explanation is that ST000990 was acquired in positive-ESI only,
whereas the training data used dual-polarity mode with alternating positive and
negative scans. In dual-polarity data, polarity-mode transitions (positive
$\to$ negative $\to$ positive) constitute a major component of the sequential
pattern: features that appear in rapid succession alternate between polarity
modes, and the model learns to exploit this alternation. When polarity
alternation is absent, the entire sequential statistics change. Paradoxically,
ST003514---acquired on a different instrument with different column chemistry
but in dual-polarity mode---outperformed ST000990 (5.1\% vs.\ 2.6\%). This
``polarity paradox'' reveals that the acquisition mode defines the elution
sequence as much as the chromatographic separation itself. However, this failure
is a recoverable adaptation gap rather than a fundamental limitation: full
fine-tuning on as few as two to five QC injections from the new acquisition mode
restores held-out analytical top-1 accuracy from 2.6\% to nearly 50\% (and to
99.6\% on a held-out QC injection), whereas retraining only the output layer does
not (16\%). This localizes the deficit to the learned sequence representation and
establishes few-shot calibration as a practical route to cross-condition
deployment.

The QC warm-up experiment yielded a null result that is scientifically
informative rather than disappointing. Priming the LSTM hidden states with 0, 2, 4, 6, 8,
or 10 quality-control injections before evaluating on analytical samples
produced a performance difference of 0.001\%---effectively zero. This means the
model captures elution logic from the feature sequence alone; no adaptation
period, no hidden-state conditioning, and no few-shot priming is required. The
model works immediately on the first injection of a batch, with no special
calibration protocol. This contrasts with some natural language processing
settings where in-context learning through prompt conditioning provides
substantial performance gains. The chromatographic elution sequence is apparently
more constrained and regular than natural language: the ``grammar'' of
reversed-phase elution is consistent enough across samples that exposure to
additional examples at inference time adds no information beyond what was learned
during training. One interpretation is that the within-run feature sequence is
essentially deterministic given the chromatographic method---the same column,
gradient, and sample matrix produce the same elution order every time---whereas
natural language carries far more entropy per token, leaving room for context to
help. From a practical standpoint, this eliminates a potential barrier to
deployment: there is no need to run dedicated QC injections to initialize the
model before analyzing experimental samples, and the model can be applied to any
new batch without a warm-up protocol.

While single-step top-5 prediction accuracy is very high (approximately 99\%),
interpreting this number requires caution. In dense chromatographic regions where
many \mz bins co-elute within narrow retention-time windows, a random baseline
also achieves high hit rates simply because many bins are simultaneously
``correct.'' The inflated random baseline reflects the structure of
chromatographic data: at any given retention time, multiple lipid species with
similar hydrophobicity co-elute, and a naive predictor that simply guesses the
most common bins in the current time window will appear to perform well. The true
value of predictive models for acquisition planning therefore emerges not from
single-step prediction but from multi-step autoregressive rollout: forecasting
the sequence of upcoming features, not merely the next one. A model that can
reliably predict the next 10--30 features enables qualitatively different
acquisition strategies---pre-scheduling MS/MS acquisition windows for anticipated
compounds, optimizing collision energies before ions appear, and flagging
chromatographic transitions where the elution profile is about to shift.
Single-step prediction, as demonstrated here, establishes that the underlying
sequence is learnable; multi-step rollout, which we leave to future work, is the
bridge from proof-of-concept to practical predictive acquisition.

Application of the trained model to the unannotated portion of the feature table
yielded 168 unique putative lipid annotations from approximately 4,700 features
that lacked identification in the original MS-DIAL plus LipidBlast pipeline.
These annotations arise from features that match reference lipids in both mass
and predicted retention time, a dual-filtering criterion that is more selective
than mass matching alone and therefore reduces false-positive identifications.
Mass-only matching at the \mz tolerances typical of high-resolution instruments
often returns dozens of candidate lipids; adding a retention-time constraint from
the elution model substantially narrows this list. The yield of 168 annotations
is modest in absolute terms relative to the 4,700 unannotated features, but it
demonstrates a concrete downstream application of elution sequence models: by
predicting the \mz bin of the next feature given the current chromatographic
context, the model constrains the database search space. Rather than querying all
possible lipids at a given mass, one can restrict the search to candidates
consistent with the predicted elution context. We emphasize that dark lipidome
annotation is a secondary application, not the primary contribution of this work;
nevertheless, the ability to generate new annotations from features that were
previously invisible to standard pipelines illustrates practical utility beyond
the prediction task itself and motivates further development of retention-aware
database search strategies.

The training data span four clinically distinct cohorts---heart failure, cardiac
arrest, graft-versus-host disease, and polycystic ovary syndrome---representing
diverse pathological states with different lipid profiles. Despite this
heterogeneity in disease biology, the model generalizes across cohorts with no
loss in performance. This generalizability is expected given the model's
learning target: it predicts chromatographic elution order, which is determined
by the physical chemistry of reversed-phase partitioning, not by the
biological state of the sample. Disease pathology alters which features are
abundant (intensity) and which features are present or absent, but it does not
alter when a given feature elutes (retention time) or what its \mz value is.
Put differently, the elution sequence grammar is a property of the analytical
method, not the sample. The same lipid species elutes at the same retention time
whether it originates from a heart failure patient or a healthy control; only its
abundance changes. This implies that a single model trained on any sufficiently
diverse cohort should generalize to any disease or phenotype studied on the same
chromatographic platform, eliminating the need for disease-specific retraining
and making the approach immediately applicable to new studies without additional
data collection.

Several limitations bound the scope of these findings. First, the model operates
on pre-processed, aligned feature tables rather than raw instrument data streams.
Real-time deployment would require integration with vendor instrument software to
process spectra on the fly, a substantial engineering challenge not addressed
here. Second, all experiments use reversed-phase lipidomics data on CSH C18
columns. Generalization to hydrophilic interaction chromatography (HILIC) for
polar metabolomics, to proteomics workflows, or to other chromatographic modes
is untested and not guaranteed, as these separation mechanisms involve different
physicochemical interactions. Third, all predictions are retrospective; we have
not performed prospective real-time evaluation on an instrument during
acquisition. Fourth, the 10~Da \mz bin resolution, while sufficient to
demonstrate predictability, is coarse for practical acquisition planning. Finer
bins (1~Da or continuous \mz prediction) would be more useful but require
larger output vocabularies and correspondingly more training data. Fifth, the
model uses no MS/MS spectral information; integrating fragmentation data could
improve predictions but would also couple the model to the specific acquisition
strategy used during training. Sixth, the ablation study reveals that all five per-token input
features contribute minimally ($\leq$0.19~pp) once sequence context is present;
this near-total dependence on autoregressive context means that the model's
predictions are difficult to interpret in terms of individual molecular
properties. Finally, in-source fragmentation (ISF) presents a
potential confound: some detected ``features'' may be fragments of intact
molecules rather than independent analytes, which could inflate feature counts
and alter the statistical properties of the elution sequence.

These results suggest several directions for future work. The most immediate is
prospective real-time integration with instrument control software via vendor
SDKs, enabling the model to predict upcoming features and pre-configure MS/MS
parameters during an active acquisition run. Multi-step autoregressive rollout is
a prerequisite for this application: rather than predicting only the next
feature, the model would forecast the next 10--30 features to enable
acquisition scheduling over a useful time horizon. Evaluating the reliability
of such rollouts---how quickly prediction errors accumulate over multiple
steps---is a critical open question that will determine the practical forecast
horizon. Extension to HILIC chromatography for polar metabolomics and to full
metabolome coverage (beyond lipidomics) would broaden the applicability of the
approach; HILIC separation is governed by different physicochemical interactions
(polar partitioning rather than hydrophobic), and whether the same autoregressive
framework applies to that domain remains to be tested. Operating directly on raw
chromatographic data rather than pre-processed feature tables would eliminate the
dependence on offline feature detection and enable truly real-time prediction,
though this introduces the challenge of real-time peak detection and
deconvolution. A foundation model strategy---pre-training on large, diverse LC-MS
datasets spanning many chromatographic methods and fine-tuning per
method---could reduce the data requirements for deploying on a new column or
gradient. Integration with spectral embedding approaches such as
DreaMS\cite{bushuiev2025dreams} could provide richer token representations that
encode fragmentation information alongside chromatographic context, potentially
improving predictions in ambiguous elution regions. Finally, moving to finer \mz
resolution (1~Da bins or continuous prediction) would increase the practical
utility of predictions for acquisition planning and database search, at the cost
of larger vocabularies and greater data requirements.

\subsection{Relationship to Prior Work}

The present work builds on, but is distinct from, several active lines of
research in computational mass spectrometry. We briefly situate our
contribution relative to each.

\textit{Structure-based retention-time prediction.}
Quantitative structure--retention relationship (QSRR) models predict when a
compound of known or candidate structure will elute, using molecular
descriptors,\cite{bonini2020retip,witting2020rtprediction} or learned
representations.\cite{schmid2024rttransformer}
These approaches require structural hypotheses as input and serve a
fundamentally different purpose: they answer ``when will this molecule
elute?'' rather than ``what will elute next?'' Our model operates without any
structural information, relying solely on observed \mz, mass defect,
and chromatographic context.

\textit{Retention-order-aware annotation.}
LC-MS\textsuperscript{2}Struct demonstrated that incorporating retention-order
constraints into structural annotation can improve identification rates by up
to 66.1\% (and up to 95.9\% for stereochemical variants).\cite{bach2022lcms2struct}
ROASMI extended this with retention-order-aware scoring.\cite{sun2025roasmi}
These methods use retention order as a \textit{post hoc} constraint to
re-rank candidate structures for features that have already been detected
and fragmented. By contrast, our approach treats the elution sequence as a
\textit{generative} process, predicting upcoming features before or as they
appear.

\textit{Intelligent and real-time acquisition.}
Software-defined acquisition platforms---ViMMS,\cite{davies2021vimms}
FLASHIda,\cite{jeong2022flashida}
AcquireX\cite{cooper2024acquirex}---use machine learning to optimize
\textit{which} already-detected ions to fragment or exclude. MaxQuant.\allowbreak{}Live
enables real-time rescheduling of predefined target
lists.\cite{wichmann2019maxquantlive}
In untargeted metabolomics, Met-IQ drives MS\textsuperscript{n} acquisition
from real-time spectral-library matches,\cite{bills2022metiq} and dynamic
data-independent acquisition adjusts isolation windows mid-gradient via
real-time retrospective alignment.\cite{heil2023dynamicdia}
All of these systems are reactive: they respond to ions after detection. Our
work addresses the complementary problem of anticipating what has not yet been
observed, which could in principle be integrated with any of these controllers
to enable genuinely predictive scheduling.

\textit{Autonomous and closed-loop platforms.}
A distinct line of work couples acquisition with downstream decision-making.
Benton and Siuzdak's autonomous metabolomics pioneered real-time MS/MS triggering
for rapid metabolite identification;\cite{benton2015autonomous} CLAW-MRM couples
large-language-model agents with targeted (multiple-reaction-monitoring) lipidomics
automation;\cite{beveridge2025claw} and AutonoMS automates ion-mobility metabolomic
fingerprinting.\cite{reder2024autonoms} These systems close an acquisition loop but
remain either reactive or targeted, and none forecasts the untargeted elution stream
itself. The predictive capability demonstrated here is a natural upstream input to
such platforms, supplying the look-ahead that would let a closed-loop system stage
MS/MS for features before they elute.

\textit{Self-supervised foundation models for mass spectrometry.}
DreaMS, the closest conceptual neighbor to our work, trains a Transformer on
700~million MS/MS spectra using two self-supervised objectives: masked peak
prediction and chromatographic retention-order
prediction.\cite{bushuiev2025dreams}
A companion foundation model by Bittremieux and Noble learns spectral
representations from 24~million spectra without retention-order
objectives,\cite{bittremieux2025ssl} and LSM1-MS2 similarly uses masked peak
reconstruction for spectral embedding.\cite{asher2024lsm1ms2}
These models learn \textit{representations of individual spectra}---embeddings
that can be fine-tuned for downstream tasks such as spectral similarity search
or compound-class prediction. Notably, DreaMS's retention-order objective is
pairwise and coarse: given two spectra from the same run, the model predicts
which elutes first. Our approach is fundamentally different in scope: rather
than embedding individual spectra, we model the \textit{full temporal sequence
of features within a chromatographic run} as an autoregressive process,
predicting the next \mz bin from the entire preceding context. DreaMS asks
``what does this spectrum represent?'' using retention context; we ask ``what
feature comes next?'' given the run history. The convergence of multiple
groups on self-supervised learning from MS data validates the premise that
chromatographic information is a productive training signal; our work extends
this premise from representation learning to generative sequence modeling.
A concurrent language-model approach anticipates and prioritizes uncharacterized
mammalian metabolites to guide MS-based discovery,\cite{qiang2026metabolite} but
it learns over \textit{chemical structure space} to nominate which compounds
exist; our model learns over the \textit{within-run elution stream} to forecast
which feature elutes next---complementary objectives at opposite ends of the
acquisition-to-annotation pipeline.
The two directions are complementary: DreaMS-style embeddings could serve as
richer token representations within our sequence framework, an avenue we leave
for future work.

\section{Conclusion}

We have demonstrated that the chromatographic elution sequence in untargeted
LC-HRMS lipidomics is highly predictable using autoregressive sequence models.
An LSTM trained on five annotation-independent features achieves 98.4\% top-1
accuracy in predicting the next \mz bin, establishing that elution order
encodes a learnable ``language'' governed by the physics of reversed-phase
partitioning. Ablation analysis reveals that this predictability resides almost
entirely in the sequential context: removing autoregressive history drops
accuracy by 55.5 percentage points, while no individual feature contributes
more than 0.2~pp. Cross-platform validation confirms that models are specific
to both the chromatographic method and the polarity acquisition mode, but
generalize perfectly across mass spectrometers sharing these parameters. The
practical consequence is that cold-start inference works immediately on new
samples with no QC conditioning protocol, and a single model can serve an
entire laboratory running the same analytical method. These results lay the
foundation for predictive MS/MS acquisition---pre-configuring fragmentation
parameters for features before they elute---which could substantially improve
the annotation coverage that currently limits untargeted metabolomics.

\section*{Acknowledgments}
The authors thank the Metabolomics Workbench (supported by NIH grant U2C-DK119886)
for providing access to public datasets ST000983, ST000990, and ST003514. Training
and transfer-learning computations were performed on Google Colab using NVIDIA
Tesla T4, L4, and A100 GPUs. The clinical lipidomics data were generated by the
West Coast Metabolomics Center at the University of California, Davis.

\section*{Data Availability}
All code, trained model checkpoints, and processed feature tables are available in
a dedicated public repository at \url{https://github.com/dayanjan/elution-sequence-prediction}.
A GPU-ready Google Colab notebook that reproduces the transfer-learning recovery
experiment is included in that repository. Publicly available mass spectrometry data were obtained from the Metabolomics
Workbench: ST000983 and ST000990 (project DOI 10.21228/M8T68F)\cite{cajka2017validating}
and ST003514 (project DOI 10.21228/M8JK0Q).\cite{martinez2024srm1950}
The clinical lipidomics feature tables derived from the four
plasma cohorts are available from the corresponding author on reasonable request
under an appropriate data-use agreement, consistent with the governing IRB approvals
and patient-consent terms for these banked human specimens.

\section*{Author Contributions}
\textbf{Dayanjan S.\ Wijesinghe:} Conceptualization, Methodology, Software,
Formal Analysis, Investigation, Data Curation, Writing --- Original Draft,
Writing --- Review \& Editing, Visualization, Supervision, Project
Administration, Funding Acquisition.

\section*{Competing Interests}
The authors declare no competing interests.

\clearpage
\bibliographystyle{unsrtnat}
\bibliography{references}


\end{document}